\documentclass[12pt]{article}
\usepackage{amsfonts,amsmath,amssymb,mathrsfs,url}

\title{Bohmian Mechanics at Space-Time Singularities.\\
   II. Spacelike Singularities}

\author{
Roderich Tumulka\footnote{Department of Mathematics, 
     Rutgers University, 110 Frelinghuysen Road, Piscataway, NJ 08854-8019, USA.
     E-mail: tumulka@math.rutgers.edu}
}
\date{May 20, 2009}

\newcommand{\Hilbert}{\mathscr{H}}

\newcommand{\conf}{\mathcal{Q}}
\renewcommand{\Re}{\mathrm{Re}}
\renewcommand{\Im}{\mathrm{Im}}

\newcommand{\PPP}{\mathbb{P}}
\newcommand{\RRR}{\mathbb{R}}
\newcommand{\CCC}{\mathbb{C}}

\newcommand{\SSS}{\mathbb{S}}

\newcommand{\scp}[2]{\langle #1|#2 \rangle}
\newcommand{\pr}[1]{| #1 \rangle \langle #1 |}

\newcommand{\Laplace}{\Delta} 
\DeclareMathOperator{\tr}{tr}
\DeclareMathOperator{\diag}{diag}
\DeclareMathOperator{\grad}{grad}

\newcommand{\st}{\mathscr{M}} 
\newcommand{\sing}{\mathscr{S}} 
\newcommand{\surface}{\tilde{\sing}} 
\newcommand{\foliation}{\mathcal{F}} 
\newcommand{\spin}{S} 
\newcommand{\fer}{\mathrm{f}}
\newcommand{\bos}{\mathrm{b}}
\newcommand{\eith}{\mathrm{x}}
\newcommand{\dens}{p} 
\newcommand{\dm}{\hat\rho} 
\newcommand{\Lop}{\mathcal{L}} 
\newcommand{\superop}{\mathcal{S}} 
\newcommand{\aq}{r} 
\newcommand{\bq}{q'} 
\newcommand{\cq}{\tilde{q}} 

\newcommand{\be}{\begin{equation}}
\newcommand{\ee}{\end{equation}}

\begin{document}
\maketitle
\begin{abstract}
We develop an extension of Bohmian mechanics by defining Bohm-like trajectories for quantum particles in a curved background space-time containing a spacelike singularity. As an example of such a metric we use the Schwarzschild metric, which contains two spacelike singularities, one in the past and one in the future. Since the particle world lines are everywhere timelike or lightlike, particles can be annihilated but not created at a future spacelike singularity, and created but not annihilated at a past spacelike singularity. It is argued that in the presence of future (past) spacelike singularities, there is a unique natural Bohm-like evolution law directed to the future (past). This law differs from the one in non-singular space-times mainly in two ways: it involves Fock space since the particle number is not conserved, and the wave function is replaced by a density matrix. In particular, we determine the evolution equation for the density matrix, a pure-to-mixed evolution equation of a quasi-Lindblad form. We have to leave open whether a curvature cut-off needs to be introduced for this equation to be well defined.

\medskip

  \noindent PACS numbers:
  04.20.Dw; 
  03.65.Ta; 
  04.62.+v. 
  Key words: 
  quantum theory in curved background space-time; 
  Schwarzschild space-time geometry;
  spacelike singularity; 
  Bohmian trajectories; 
  particle creation and annihilation;
  pure-to-mixed evolution.
\end{abstract}
\tableofcontents

\section{Introduction}

This paper, part two of a two-part series on Bohmian mechanics at space-time singularities, can be read independently of part one \cite{Tum07}. We consider quantum mechanics in a relativistic space-time with fixed background metric containing spacelike singularities from the perspective of Bohmian mechanics (also known as pilot-wave theory), a precise version of quantum mechanics in which particles have world lines. I argue that, in this setting, a unitary time evolution for the wave function is no longer possible, and must be replaced by an equation for a density matrix, see \eqref{dmevol} below, an integro-differential equation of a quasi-Lindblad form that evolves pure states into mixed states. To my knowledge, this equation is novel; but it is in line with an earlier proposal of Hawking \cite{Haw76,Haw82}, grounded on black hole evaporation, to the effect that the fundamental physical evolution law should transform pure states into mixed states.

The role of the density matrix here is unusual: Usually, density matrices represent statistical mixtures, or, in the case of a reduced density matrix obtained by a partial trace, the state description of a system that is entangled with its environment. Here, in contrast, the density matrix does not represent incomplete knowledge but rather determines the motion of the particles, a role normally played in Bohmian mechanics by the \emph{wave function}. Still, the evolution involves \emph{information loss}, as different density matrices at one coordinate time may evolve into the same density matrix at a later time.

According to the singularity theorems of general relativity \cite[chap.~8]{HE73}, a black hole arising from a gravitational collapse contains a singularity, which is generally believed to be spacelike. As a concrete example of our general scheme we consider $N$ non-interacting spin-$\tfrac{1}{2}$ particles in a Schwarzschild space-time, which contains two spacelike singularities. The terminology and notation we use is that of quantum mechanics, rather than quantum field theory. The evolution equation for the density matrix then contains the Dirac Hamiltonian and a term connected to the singular boundary of the configuration space, while the configuration space arises from a spacelike hypersurface bordering on the singularity. Since the model assumes that the space-time metric is given, it does not include any back reaction such as growth of the horizon or the singularity after swallowing particles.

A crucial fact for the development of our Bohm-type model is that the particle world lines are everywhere timelike or lightlike and thus can begin but not end on a past spacelike singularity (hereafter, \emph{past singularity}) and end but not begin on a future spacelike singularity (hereafter, \emph{future singularity}). That is why the discussion of spacelike singularities is very different from that of timelike singularities. In the absence of other mechanisms of particle creation and annihilation, the number of particles can only decrease (increase) in the presence of a future (past) singularity. Since the particle number is not constant, we are forced to use \emph{Fock space}, usually used only in quantum field theory but not in quantum mechanics. The natural laws for the Bohmian particles specify the particles' velocities and assert that any particle that hits a future singularity disappears while the other particles continue moving along the appropriate Bohm-like trajectories. Instead of merely adding Bohmian trajectories to known ``orthodox'' quantum theories, we use Bohmian mechanics to find the appropriate evolution equation for the density matrix.

The quasi-Lindblad equation that we obtain for the density matrix arises also in a different context, replacing the singularity $\sing$ by a spacelike hypersurface $\surface$: it arises from a unitarily evolving wave function $\psi_t$ by tracing out those degrees of freedom localized in the future of $\surface$. In this scenario, the density matrix does represent only partial information about the true quantum state, and the quasi-Lindblad equation represents the procedure of \emph{continuously tracing out} (continuously in time) more and more degrees of freedom (corresponding to larger and larger portions of space). 

It might seem that there is the following alternative to our quasi-Lindblad equation: We may refuse to replace the wave function $\psi_t$ with a density matrix, stick to the Dirac equation for $\psi_t$, and just accept that the time evolution is not unitary. This amounts to deleting the amount of wave function that has crossed the singular boundary of configuration space. The fact that $\|\psi_t\|$ will shrink with increasing $t$ may seem natural as $\|\psi_t\|^2$ represents the probability that no particle has hit the singularity up to time $t$. But what that really means becomes clear from the Bohmian point of view: it means that when \emph{one} particle hits the singularity, \emph{all} particles disappear. And that is a much less natural dynamics than postulating that when one particle hits the singularity, all other particles continue moving along Bohm-type trajectories.

Due to limits of my knowledge, I have to leave open the question whether a \emph{curvature cut-off} needs to be introduced to make the quasi-Lindblad equation well defined; I conjecture that no such cut-off is necessary. Such a cut-off can be implemented by choosing an arbitrarily thin neighborhood of the singularity $\sing$ whose surface is a spacelike hypersurface $\surface$, and let $\surface$ play the role of $\sing$ in the quasi-Lindblad equation. It seems that this question boils down to the question whether the probability current associated with a solution of the 1-particle Dirac equation possesses a continuation on the singular boundary $\sing$; I conjecture that the answer is yes. 
Furthermore, for the precise formulation of the quasi-Lindblad equation it is relevant whether the spin spaces can be defined on the singular boundary $\sing$ in such a way that also a wave function obeying the 1-particle Dirac equation possesses a continuation on $\sing$. In this paper, however, we will simply assume that limits on $\sing$ exist whenever needed.

Further questions arise from past singularities as in the \emph{white hole} of the Schwarzschild space-time. By time reversal symmetry, the same evolution equation for the density matrix that holds at a future singularity should apply here backwards in time. When looked at in the ordinary time direction, from the past to the future, then the evolution of the density matrix will be indeterministic because of the ``information loss'' property in the other time direction. This kind of indeterminism is distinct from the quantum indeterminism (as expressed by the Heisenberg uncertainty relation and indeed compatible with the determinism of Bohmian mechanics) and the indeterminism represented by the stochastic law governing the configuration in Bell-type quantum field theories \cite{crlet,crea2B}. Once the evolution of the density matrix is fixed, the evolution law for the configuration is fixed by time reversal symmetry, and turns out to involve, besides a Bohm-type equation of motion, a stochastic law for particle creation at the singularity.

Another work on Bohmian mechanics and black holes is Valentini's \cite{Val04}, proposing that the equivariance of the $|\psi|^2$ distribution might be violated in the presence of black holes.

\bigskip

This paper is organized as follows. In the remainder of this section we give an overview. In Section~\ref{sec:Bohm} we recall Bohm's law of motion for the Dirac equation. In Section~\ref{sec:schwarz} we recall the Schwarzschild metric. In Section~\ref{sec:future} we develop our new version of Bohmian mechanics in the presence of a future singularity, and in Section~\ref{sec:past} in the presence of a past singularity.

\subsection{Horizons}

Let me make a few remarks about the status of horizons in Bohmian mechanics.

Consider a many-particle quantum system in a background space-time containing a black hole. From the point of view of orthodox quantum mechanics, it is natural to trace out all degrees of freedom inside the event horizon. From the Bohmian viewpoint, in contrast, this is not natural. Instead, it is natural to ask what happens behind the horizon. This is so because from the orthodox viewpoint, the most important question is what an observer will see, and it is a frequent assumption that the relevant observers sit at infinity. From the Bohmian viewpoint, the most important question is what actually happens. Thus, to define Bohmian mechanics in a curved space-time, we need to define as well the trajectories inside the black hole. (A viewpoint that, like the ``Copenhagen'' view of quantum mechanics, dismisses any theory of particle positions because it regards the latter as ``hidden variables,'' may naturally tend to dismiss as unreal also everything that is hidden behind a horizon.)

Thus, the Bohmian viewpoint leads us to the following attitude:
What happens inside a black hole \emph{can and should} be described by a physical theory.

Indeed, by the nonlocal character of Bohmian mechanics, the velocities of the particles outside the black hole may depend on the positions of the particles inside the black hole.  But all this requires no additional research for the definition of the theory, as the Bohm-type law of motion we use (see Eq.~\eqref{hbd} below) automatically implies influences across event horizons. What requires further work is not the presence of a horizon, but the presence of a singularity. (The need for this further work would disappear if none of the hypersurfaces belonging to the time foliation bordered on the singularity. However, for our purposes that is the uninteresting case. Furthermore, I see no reason why the time foliation should avoid the singularity.)

Now we can appreciate the differences between the approach of this paper and that of Hawking \cite{Haw76}: Hawking traces out what has passed the \emph{horizon}, while we trace out what has hit the \emph{singularity}; for Hawking, taking the partial trace is only a matter of representing the knowledge of observers at infinity, while for us it is part of defining the true particle trajectories; Hawking uses positivist arguments, while we start from a realist picture; for Hawking, a pure-to-mixed evolution fundamentally occurs only when the black hole evaporates, while for us it occurs continuously during the existence of the singularity; for Hawking, the pure-to-mixed evolution occurs because late (i.e., post-evaporation) observers cannot measure early (i.e., pre-evaporation) observables, while for us even a demon knowing all facts has to apply the pure-to-mixed evolution; Hawking focuses on the quantum state at infinity (as in scattering theory), while we need the full time evolution.

\subsection{Background}

\emph{Bohmian mechanics} was developed as a realist version of nonrelativistic quantum mechanics \cite{Bohm52} and succeeds in explaining all phenomena of quantum mechanics in terms of an objective, observer-independent reality consisting of point particles moving in space; see \cite{Gol01} for an overview. Bohmian mechanics possesses a natural extension to relativistic space-time if a preferred foliation of space-time into spacelike hypersurfaces (called the \emph{time foliation} $\foliation$) is granted \cite{HBD}. This extension has also been formulated for curved space-time geometries \cite{3forms,Tum06d}, but not yet for geometries with singularities. While horizons present no difficulty, singularities require further work to define the theory: Basically, we have to specify what happens when a particle hits the singularity, since at this point the law of motion is no longer defined. The possibility we consider here is that the particle gets annihilated: that is, if the system consisted of $N$ particles, then the world line of the particle that has arrived at the singularity ends there, while the other $N-1$ particles, which are not at the singularity and thus have no reason to vanish, continue to move according to Bohm's law of motion. To make this possible, we need wave functions from Fock space, i.e., superpositions of different particle numbers, as always when particles can get created or annihilated. Further considerations then naturally lead us to specific equations, defining a Bohm-type theory.

\bigskip

\emph{Space-time singularities} are points on the boundary of space-time where the metric cannot be extended smoothly because the curvature becomes infinite. There is no universally accepted mathematical construction of these boundary points from a given metric (see chapter 8 of \cite{HE73} for a discussion), but readers may adopt the construction of \cite{GKP72}, called the \emph{causal boundary}, that defines boundary points as terminal indecomposable past sets (TIPs) or terminal indecomposable future sets (TIFs), together with suitable identifications. A boundary point that does not lie at infinity is considered a singular boundary point. The singular boundary is \emph{timelike} at a point $x$ if $x$ has a non-empty TIP \emph{and} a non-empty TIF; it is \emph{future spacelike} if $x$ has empty TIF and \emph{past spacelike} if it has empty TIP. Like a hypersurface, a singularity can be timelike at some points and spacelike at others. 

We will not construct the singular boundary but regard it as given. More precisely, we assume that space-time $\st$ is a \emph{manifold with boundary} \cite{Lan72,BM}, where the boundary represents the singularity. In particular, we assume that the singularity has the structure of a 3-dimensional manifold. In the Schwarzschild geometry, for example, $\st$ is diffeomorphic to $[-1,1]\times\RRR\times\SSS^2$; the singular boundary is everywhere spacelike and has two connected components $\sing_1, \sing_2$, each of topology $\RRR\times \SSS^2$; one is a future singularity, the other a past singularity. We shall recall some details about the Schwarzschild geometry in Section~\ref{sec:schwarz}.

\subsection{Motivation}\label{sec:motivation}

It is a natural part of the research program on Bohmian mechanics to extend the theory to more general quantum theories, to all kinds of settings. To the extent that we have reason to believe that singularities exist in our universe, we obtain here a more appropriate version of Bohmian mechanics. Concerning techniques of constructing Bohm--type models, we find that the Bohm-type law of motion proposed by D\"urr \emph{et al.}~\cite{HBD} for relativistic space-time with a foliation works unproblematically even in the presence of a singularity, a result that lends support to this law of motion.  

Since Bohmian mechanics is a particularly precise and unambiguous version of quantum mechanics it may serve as a tool for studying quantum mechanics in curved space-time. Thus, our study can as well be regarded as one on \emph{quantum mechanics at space-time singularities}. 

It is remarkable that the presence of spacelike singularities forces us to change the structure of Bohmian (or quantum) mechanics in three ways: First, as a consequence of the non-conservation of particle number, it is necessary to use Fock space; second, a density matrix replaces the wave function in its role of guiding the particles; third, the evolution is no longer unitary.

Independently of the Bohmian viewpoint, we obtain a novel evolution equation for the density matrix in the presence of a spacelike singularity.  It can be argued that this evolution of the density matrix is more fundamental than the unitary dynamics of 
quantum mechanics, since if our universe contains any spacelike 
singularities then a unitary evolution apparently does not exist. Conversely, if we think that the fundamental evolution should be unitary, we should be skeptical about the existence of spacelike singularities. In contrast, timelike singularities do not enforce deviations from unitarity and are thus less dramatic in Bohmian (or quantum) mechanics than spacelike ones; this situation differs from that in classical mechanics, where future spacelike singularities are unproblematic while timelike and past spacelike singularities are not covered by the laws of classical mechanics---anything could come out of such singularities.

When using a foliation that consists of Cauchy hypersurfaces (which, in particular, do not border on the singularity), the usual unitary evolution works and no need for a quasi-Lindblad evolution arises. So why even consider a non-Cauchy foliation? One reason is that for a non-Cauchy foliation, something unusual and interesting happens: the unitary evolution gets replaced by the quasi-Lindblad evolution. A second and more serious reason is that, according to Bohmian mechanics (in the version that we use here), there is a fact in nature about which foliation is the right one, i.e., which is the time foliation $\foliation$. Since the law selecting $\foliation$ is presently not known, we have to admit the possibility that $\foliation$ is a non-Cauchy foliation. Indeed, a candidate for a law for $\foliation$ is formulated in \eqref{lawF} below, and this law leads inevitably to non-Cauchy foliations in the presence of a singularity. This observation urges us to consider non-Cauchy foliations.

\subsection{Overview of New Equations}
\label{sec:overviewneweq}

Let us have a first look at relevant equations, postponing a more careful definition and deeper discussion to Section~\ref{sec:future}. Let $\sing$ be a future singularity, regarded as a 3-dimensional manifold, the boundary of space-time.

The evolution equation for the density matrix $\dm_t$ on (either bosonic or fermionic) Fock space is an integro-differential equation and reads
\be\label{dmevol}
\frac{\partial\dm_t}{\partial t} 
=\tfrac{i}{\hbar} [\dm_t,\hat H_t] + \Lop \dm_t\,,
\ee
with $\hat H_t$ the Hamiltonian on Fock space arising from the Dirac equation; the square brackets mean the commutator. The symbol $\Lop$ denotes a particular superoperator (i.e., an operator acting on density matrices rather than wave functions) which can be defined explicitly by
\begin{align}
(\Lop \dm_t)(q;\aq) &= \sqrt{(\#q+1) (\#\aq+1)} \nonumber
\int\limits_{\sing_t} d^2x  \: w(x) \:\times\\
&\quad \times \lim_{y\to x, y\notin \sing} 
c_\mu(x) \, d_4(y)\:
\tr_{\spin_{y}}\Bigl(\dm_t(q,y;\aq,y) \: \alpha^\mu(y) \Bigr)\,.
\label{Lop}
\end{align}
Let us explain the notation. Let $\Sigma_t$ be the spacelike hypersurface corresponding to time $t$ and $\spin_y$ the spin space at space-time point $y$; with a configuration $q=(q_1,\ldots,q_N)\in \Sigma_t^N$ there is associated the spin space
\be\label{spinqdef}
\spin_q = \spin_{q_1}\otimes \cdots \otimes \spin_{q_N}\,.
\ee
We write density matrices $\dm$ as spinor-valued functions $\dm(q;\aq)$ of two configuration variables, corresponding to the formal expression
\be
\dm(q;\aq)= \scp{q}{\dm|\aq}\,,
\ee
where the scalar product is a partial scalar product taken only over the position degrees of freedom but not the spin degrees of freedom; $\dm(q;\aq)$ is a linear mapping $\spin_{\aq}\to\spin_q$. The symbol $\#q$ denotes the number of particles in the configuration $q$. Let $\sing_t=\sing\cap \Sigma_t$, $n^\mu(x)$ the unit normal vector on $\Sigma_t$ at $x\in\Sigma_t$, and
\be\label{alphadef}
\alpha^\mu(x) = \bigl(n_\nu(x)\,\gamma^\nu(x)\bigr)^{-1}\, \gamma^\mu(x)\,,
\ee
where $\gamma^\mu$ are the same gamma matrices as in the Dirac equation.
Furthermore, in \eqref{Lop} $d^2x$ denotes the coordinate area measure of the surface element; $w(x)\, dt$ is the thickness in coordinate space of the 3-dimensional strip between the 2-dimensional surfaces $\sing_t$ and $\sing_{t+dt}$ at $x\in \sing_t$; $c_\mu(x)$ is the 4-vector which \emph{in coordinate space} is orthogonal to $\sing$ at $x$, points to the future, and has unit length; $d_4$ is the determinantal 4-volume factor,
\be\label{d4def}
d_4(y) = \sqrt{-\det g_{\mu\nu}(y)}\,;
\ee
$\tr_{\spin_y}$ denotes the partial trace over $\spin_y$; and the pair $q,y$ means the configuration arising from $q$ by adding a particle at $y$.\footnote{Although some quantities in \eqref{Lop} explicitly refer to a coordinate system, the formula \eqref{Lop} is actually equally valid in every coordinate system that has the $t$ function as its time coordinate. 
Concerning the limit $y\to x, y \notin\sing$, I conjecture that the spin bundle can be extended to the singularity in such a way that $\dm_t(q,x;\aq,x)$ is well defined for $x\in \sing$ without the need of a limit, and that $\displaystyle\lim_{y\to x, y\notin \sing} c_\mu(x) \, d_4(y) \, \alpha^\mu(y)$ is a well-defined endomorphism $\alpha_\sing(x)$ of $\spin_x$.}

The Bohm-type trajectories are defined by a first-order equation for the particle configuration $Q(t) = (Q_1(t),\ldots,Q_N(t))\in \Sigma_t^N$,
\be\label{Bohmdm}
\frac{dQ_k^{\mu}}{dt} = 
\frac{d_4(Q_k)}{d_3(Q_k)}\, 
\frac{\tr_{\spin_q} \bigl( \dm(q;q) \alpha_k^\mu(q) \bigr)}
{\tr_{\spin_q} \dm(q;q)}\Big|_{q=Q(t)}
\ee
where $d_3$ is the determinantal 3-volume factor
\be\label{d3def1}
d_3(x) = \sqrt{-\det g^{(3)}(x)}
\ee
and 
\be\label{alphamuk}
\alpha^\mu_k(q)=I^{\otimes (k-1)}\otimes \alpha^\mu(q_k) \otimes I^{\otimes (N-k-1)}\,,
\ee
for which we will often simply write $\alpha^\mu(q_k)$, as the argument $q_k$ makes it unambiguous on which spin index the matrix acts.

In addition, it is postulated that whenever the configuration $Q(t)$ reaches the boundary of configuration space, which happens when one of the particles reaches the singularity, then that particle will be annihilated, corresponding to a jump 
\be\label{Qjump}
Q(t-) \to Q(t+) = Q(t-)\setminus \sing
\ee
in the configuration, where $Q(t\pm) = \lim_{\varepsilon\searrow 0} Q(t\pm\varepsilon)$.

The equation of motion \eqref{Bohmdm} is of the form
\be\label{Bohmj}
\frac{dQ(t)}{dt} = \frac{j(Q(t))}{\dens(Q(t))}
\ee
with the probability density $\dens$ (relative to coordinate volume) given by
\be\label{densdm1}
\dens(q) = \dens^{\dm_t}(q) = d_3(q)\,
\tr_{\spin_q} \dm_t(q;q)\,,
\ee
where
\be\label{d3def2}
d_3(q) = \prod_{k=1}^{\#q}\sqrt{-\det g^{(3)}(q_k)}\,,
\ee
and the probability current density $j$ given by
\be\label{jdm1}
j_k^\mu(q) = j^{\dm_t,\mu}_k(q) = 
d_3(q\setminus q_k)\,d_4(q_k)\,
\tr_{\spin_q} \bigl( \dm(q;q) \alpha^\mu(q_k) \bigr)\,.
\ee
Indeed, \eqref{Bohmdm} is the straightforward generalization of Bohm's equation of motion from wave functions to density matrices; by setting $\dm=\pr{\psi}$ we obtain the usual form of Bohm's equation of motion for the Dirac equation,
\be\label{BohmDirac}
\frac{dQ_k^\mu}{dt} = 
\frac{d_4(Q_k)}{d_3(Q_k)}\, 
\frac{\psi^*(Q) \alpha^\mu(Q_k) \psi(Q)}
{\psi^*(Q) \psi(Q)}
\ee
with $\phi^*\psi=\overline{\phi}\gamma^0\psi$ the inner product in spin space. The version \eqref{Bohmdm} with a density matrix has been considered before in two places: 
\begin{itemize}
\item in \cite{Belldm} for contrasting the trajectories arising from \eqref{Bohmdm} with those arising from \eqref{BohmDirac} for a random $\psi$ whose distribution has density matrix $\dm$, illustrating that in Bohmian mechanics nature needs to know $\psi$, not just $\dm$; 
\item in \cite{dm} for finding a replacement of the \emph{conditional wave function} for particles with spin.
\end{itemize}

Since, according to \eqref{Bohmdm}, the density matrix $\dm$ has the role of ``guiding'' the particles, to determine how they move, it is a \emph{fundamental} density matrix (in the terminology of \cite{dm}), as opposed to a description with less than full information. Eq.~\eqref{dmevol} defines a ``pure-to-mixed'' evolution, where a density matrix is called ``pure'' if it is a one-dimensional projection and ``mixed'' otherwise. Readers should not understand ``mixed'' as referring to any ensemble of wave functions, or as indicating the presence of any randomness. The density matrix plays not a \emph{statistical} role here, but rather the role of a \emph{fundamental} object in the theory; this difference is clear from the Bohmian perspective but hard to discern from the perspective of orthodox quantum theory.

\bigskip

The evolution equation \eqref{dmevol} is of a \emph{quasi-Lindblad} form, as we explain now. The Lindblad equation reads \cite{Lin76,GKS}
\be\label{Lin1}
\frac{\partial \dm_t}{\partial t} = \tfrac{i}{\hbar} [\dm_t,\hat H]  +
\sum_\lambda \hat A_\lambda\, \dm_t\, \hat A^*_\lambda - \tfrac{1}{2} 
\Bigl\{\dm_t,\sum_\lambda \hat A^*_\lambda \hat A_\lambda \Bigr\}\,,
\ee
where $A_\lambda$ is a countable family of bounded operators and $\hat H$ a bounded self-adjoint operator, and the curly brackets mean the anti-commutator. Every (uniformly continuous) \emph{quantum dynamical semigroup} is generated by an equation of this form \cite{Lin76,GKS}. A more general version reads
\be\label{Lin2}
\frac{\partial \dm_t}{\partial t} = \tfrac{i}{\hbar} [\dm_t,\hat H_t]  +
\int \mu_t(d\lambda) \, \hat A_t(\lambda)\, \dm_t\, \hat A_t^*(\lambda) - \tfrac{1}{2} 
\Bigl\{\dm_t,\int \mu_t(d\lambda) \hat A_t^*(\lambda) \hat A_t(\lambda)\Bigr\}\,,
\ee
where $\lambda$ is any parameter, $\mu_t$ any measure over that parameter, $\hat A_t(\lambda)$ any operator, and $\hat H_t$ a self-adjoint operator. 

Our equation \eqref{dmevol} is similar but not exactly of this type. Rather, it is of the form
\be\label{qLin}
\frac{\partial \dm_t}{\partial t} = \tfrac{i}{\hbar} [\dm_t,\hat H_t]  +
\int \mu_t(d\lambda) \, \hat A_t(\lambda)\, \dm_t\, \hat A_t^*(\lambda) \,,
\ee
and that is why we call it a quasi-Lindblad equation. There are two main differences to \eqref{Lin2}: First, our Hamiltonian $\hat H_t$ is not self-adjoint, a fact corresponding to the loss of probability at the singular boundary of configuration space. Second, the third term on the right hand side of \eqref{Lin2} is missing in \eqref{qLin}. To understand why, note that that term is needed in \eqref{Lin2} to compensate the trace of the second term: if the trace of $\dm_t$ is to be conserved, the trace of the right hand side of \eqref{Lin2} should vanish; a commutator $[\dm_t,\hat H_t]$ with a self-adjoint $\hat H_t$ has zero trace; since the second and the third term contain the same factors cyclically permuted, they have equal trace---but opposite signs in front of them. In contrast, the trace of the commutator in \eqref{qLin}, or \eqref{dmevol}, is \emph{not} zero, but instead minus the rate of probability loss at the boundary of configuration space. It is equally large, but with opposite sign, as $\tr\,\Lop\dm_t$, so that also the right hand side of \eqref{qLin} is traceless. 

In order to show that \eqref{dmevol} is of the form \eqref{qLin} we need more details about how the Hilbert space is defined; we postpone this to Section~\ref{sec:framework}, see equation \eqref{LinA}. 

For $t_2>t_1$, the mapping $\dm_{t_1} \mapsto \dm_{t_2}$ defined by solving \eqref{dmevol} can be regarded as a superoperator $\superop_{t_1}^{t_2}$ from the trace class of $\Hilbert_{t_1}$ to the trace class of $\Hilbert_{t_2}$. The superoperators are completely positive\footnote{A linear mapping from the trace class of $\Hilbert_{1}$ to the trace class of $\Hilbert_{2}$ is called \emph{completely positive} if, first, it maps positive operators on $\Hilbert_1$ to positive operators on $\Hilbert_2$ and, second, its obvious extension maps positive operators on $\Hilbert_1\otimes \Hilbert_3$ to positive operators on $\Hilbert_2\otimes \Hilbert_3$, for any $\Hilbert_3$ \cite{cpm}. There is wide consensus that this is the appropriate positivity property for transformations of density matrices.} and satisfy $\superop_{t_1}^{t_1}\dm = \dm$ and $\superop_{t_2}^{t_3}\superop_{t_1}^{t_2}=\superop_{t_1}^{t_3}$. This structure is similar to that of quantum dynamical semigroups, which often arise as effective evolution of reduced density matrices, for example representing decoherence. Here, of course, it is not decoherence that causes the pure-to-mixed evolution; it is not a reduced density matrix that obeys \eqref{dmevol} but the fundamental density matrix; and \eqref{dmevol} is not an effective equation but the fundamental law of nature replacing the Schr\"odinger equation.

The question arises how it can be possible that the equation \eqref{dmevol} fails to be of the Lindblad form \eqref{Lin1} or \eqref{Lin2}, in view of the Lindblad theorem \cite{Lin76,GKS} that (uniformly continuous) quantum dynamical semigroups are always generated by an equation of that form. One might suspect the reason is that the superoperators $\superop_{t_1}^{t_2}$ defined by \eqref{dmevol} do not form a semigroup, since the semigroup structure requires time-translation invariance (in the sense of $\superop_0^t = \superop_s^{t+s}$). While it is true that the $\superop_{t_1}^{t_2}$ do not \emph{usually} form a semigroup, one can devise cases that actually are time-translation invariant and for which the $\superop_{t_1}^{t_2}$ do form a semigroup (using a suitable identification of the Hilbert spaces associated with $t_1$ and $t_2$); so the semigroup assumption is not the relevant hypothesis of the Lindblad theorem that fails here. What fails is only the assumption of uniform continuity, so I conclude that this assumption is not, as one might have thought, merely a technical assumption without physical relevance, but a strong restriction excluding quasi-Lindblad equations.

\bigskip

For past singularities, the evolution of the density matrix into the future is not uniquely determined. If we fix one such evolution, we obtain a unique law for the rate at which particles are created at the singularity: The rate of creation of a particle at time $t$ in the 2-dimensional surface element $d^2x$ in $\sing\cap \Sigma_t$, given that the present configuration is $Q(t)=q$, is
\be\label{pastrate}
\sigma_t(d^2x|q)= 
\frac{w(x)\,(\#q+1)\, \lim\limits_{y\to x,y\notin \sing} c_\mu(x)\,d_4(y) \,
\tr_{\spin_{q,y}} \bigl(\dm_t(q,y;q,y)\:\alpha^\mu(y) \bigr)}
{\tr_{\spin_q} \dm_t(q;q)}
d^2x\,.
\ee
This formula is closely related to the creation rate formula of so-called \emph{Bell-type quantum field theories} \cite{crlet,crea2B,Bell86}, versions of Bohmian mechanics with particle creation and annihilation.

\section{Bohmian Mechanics}
\label{sec:Bohm}

Bohmian mechanics \cite{Bohm52,Bell87b,DGZ92,Gol01} is well understood in the realm of non-relativistic quantum mechanics, but needs further development in the directions of relativistic physics, quantum field theory, and quantum gravity. This paper concerns the relativistic extension in a classical gravitational field, but connects also with quantum field theory.

Bohmian mechanics postulates that particles have trajectories, governed by an equation of motion of the type
\begin{equation}\label{Bohm}
  \frac{dQ_t}{dt} = \frac{j^\psi(Q_t)}{\dens^\psi(Q_t)}\,,
\end{equation}
where $Q_t$ is the position of the particle at time $t$ (or, for a system of several particles, the \emph{configuration}), and $j^\psi$ and $\dens^\psi$ are, respectively, the quantum mechanical probability current and probability density as determined by the wave function $\psi$. 
As a consequence of the structure \eqref{Bohm} of the law of motion, if at any time $t$ the particle position (or configuration) is random with distribution $\dens^{\psi_t}$, then this is also true of any other time $t$. This property is called \emph{equivariance}. As a (quite non-obvious) consequence of \emph{that}, inhabitants of a Bohmian universe, consisting of these particles with trajectories, would observe the same probabilities in their experiments as predicted by the quantum formalism \cite{DGZ92}. That is how Bohmian mechanics explains quantum mechanics. In fact, Bohmian mechanics accounts for all phenomena of non-relativistic quantum mechanics.

\subsection{In Relativistic Space-Time}

With the invocation of a preferred foliation $\foliation$ of space-time into spacelike hypersurfaces, given by a Lorentz invariant law and called the \emph{time foliation}, it is known \cite{HBD,3forms,Tum06d} that Bohmian mechanics possesses a natural generalization to relativistic space-time. The possibility of a preferred foliation seems against the spirit of relativity, but certainly worth exploring. It is suggested by the empirical fact of quantum non-locality, and it is suggested by the structure of the Bohmian law of motion \eqref{Bohm} for many particles, in which the velocity of a particle depends on the instantaneous position of the other particles. Using a time foliation $\foliation$, a Bohm-type equation of motion was formulated in \cite{HBD} for flat space-time, and the straightforward generalization to curved space-time was formulated and mathematically studied in \cite{3forms}:
\begin{equation}\label{hbd}
  \frac{dQ_k^{\mu_k}}{ds} \propto j^{\mu_1 \ldots \mu_N} 
  \bigl(Q_1(\Sigma),\ldots, Q_N(\Sigma)\bigr) 
  \prod_{i\neq k} n_{\mu_i}\bigl(Q_i(\Sigma)\bigr)\,,
\end{equation}
where $Q_k(s)$ is the world line of particle $k\in \{1,\ldots,N\}$, $s$ is any curve parameter, $\Sigma$ is the hypersurface in $\foliation$ containing $Q_k(s)$, $n(x)$ is the unit normal vector on $\Sigma$ at $x \in \Sigma$, $Q_i(\Sigma)$ is the point where the world line of particle $i$ crosses $\Sigma$, and
\begin{equation}\label{multij}
  j^{\mu_1 \ldots \mu_N} = \overline{\psi} (\gamma^{\mu_1} \otimes \cdots \otimes
  \gamma^{\mu_N}) \psi
\end{equation}
is the probability multi-current associated with the $N$-particle Dirac wave function $\psi$. This wave function $\psi$ could either be a multi-time wave function defined on $\st^N$, where $\st$ is space-time, or, since we never use $\psi$ for configurations that are not simultaneous, it suffices that $\psi$ is defined on the $3N+1$-dimensional manifold $\bigcup_{\Sigma \in \foliation} \Sigma^N$ of simultaneous configurations. The probability density, relative to the invariant volume on $\Sigma$, is given by
\be\label{denspsi0}
j^{\mu_1 \ldots \mu_N}(q_1,\ldots,q_N) \, 
\prod_{k=1}^N n_{\mu_k}(q_k)
\ee
for $q_1,\ldots, q_N \in \Sigma$.

The extension of Bohmian mechanics to relativistic space-time that we just described does not automatically include, however, space-time geometries with singularities. The treatment of singularities requires some fundamental extensions of Bohmian mechanics, and forms a test case for the robustness of the equation of motion \eqref{hbd}. Well, the equation has stood the test, both with timelike \cite{Tum07} and spacelike singularities.

As mentioned before, the foliation might itself be dynamical. An example of a possible Lorentz invariant evolution law for the foliation is
\begin{equation}\label{lawF}
  \nabla_\mu n_\nu - \nabla_\nu n_\mu =0\,,
\end{equation}
which is equivalent to saying that the infinitesimal timelike distance between two nearby hypersurfaces from the foliation is constant along the hypersurface. This law allows to choose an initial spacelike hypersurface and then determines the foliation. A special foliation obeying \eqref{lawF} is the one consisting of the surfaces of constant timelike distance from the big bang. Note, however, that the law of motion \eqref{hbd} does not require any particular choice of law for the foliation, except that the foliation does not depend on the particle configuration (while it may depend on the wave function). Note further that in a space-time with horizons and singularities, a foliation law like \eqref{lawF} will frequently lead to hypersurfaces lying partly outside and partly inside the horizon, and indeed to hypersurfaces bordering on a singularity.

\subsection{Adapted Coordinates}

When expressed in terms of coordinates that are \emph{adapted to the time foliation}, i.e., such that the time coordinate function is constant on every time leaf $\Sigma\in\foliation$, \eqref{hbd} assumes the form \eqref{BohmDirac}, while the probability density is given by
\be\label{denspsi}
\dens(q) = d_3(q) \, \psi^*(q) \, \psi(q)\,,
\ee
and the current by
\be\label{jpsi}
j^\mu_k(q) = d_3(q\setminus q_k) \, d_4(q_k) \, \psi^*(q) \, \alpha^\mu(q_k) \, \psi(q)\,.
\ee
These equations need some elaboration.

Let us first turn to the definition of the inner product. At $x\in \st$ except on the singularity, every future-pointing timelike vector $n^\mu\in T_x \st$ defines a positive definite Hermitian inner product in the spin space $\spin_x$, usually denoted
\be
\overline\phi \, n_\mu \gamma^\mu\, \psi
\ee
for any $\phi,\psi\in \spin_x$. We will always use the inner product in $\spin_x$ defined by the future-pointing unit normal vector $n^\mu=n_\mu(x)$ on the unique time leaf $\Sigma\in\foliation$ passing through $x$, and denote that inner product by $\phi^*\psi$ for any $\phi,\psi\in\spin_x$. In terms of this inner product,
\be\label{gammaalpha}
\overline\phi \, \gamma^\mu \, \psi = \phi^* \, \alpha^\mu \, \psi
\ee
with $\alpha^\mu$ given by \eqref{alphadef}; both $\gamma$ and $\alpha$ are cross-sections of the vector bundle\footnote{We use the notation $\cup_{x\in B} E_x$, rather than $E\stackrel{\pi}{\to} B$, to denote the vector bundle over the base manifold $B$ with fiber spaces $E_x$, assuming it is clear from the context which bundle structure (as defined by the bundle maps) and, if appropriate, connection is intended.}
\be
\bigcup\limits_{x\in \st\setminus\sing} \CCC T_x\st \otimes End(\spin_x)\,,
\ee
where $\CCC T_x\st$ denotes the complexified tangent space and $End(\spin_x)$ the space of endomorphisms of $\spin_x$. From \eqref{gammaalpha} it follows that
\begin{equation}\label{multij2}
  j^{\mu_1 \ldots \mu_N} = \psi^* (\alpha^{\mu_1} \otimes \cdots \otimes
  \alpha^{\mu_N}) \psi
\end{equation}
and
\be
n_\mu(x) \alpha^\mu(x) = I\,,
\ee
where $I$ means the identity.

Expressed relative to an orthonormal basis in $T_x \st$ with $n_\mu$ as the timelike basis vector, and relative to the associated basis in spin space $\spin_x$, the gamma matrices assume their standard form; the basis in spin-space is orthonormal, $\phi^*\psi=\sum_{s=1}^4 (\phi_s)^* \psi_s$ for the components $\phi_s,\psi_s\in\CCC$ of $\phi,\psi\in\spin_x$; $\alpha^0$ is the identity; and $\alpha^1, \alpha^2, \alpha^3$ are the standard Dirac alpha matrices. 

More generally, in any coordinate system adapted to $\foliation$, $\alpha^0$ is a multiple of the identity, namely
\be\label{alpha0g00}
\alpha^0(x) = \sqrt{g^{00}(x)} \, I\,.
\ee
To see this, note that for every vector $u\in T_x\st$,
\be\label{u0nu}
u^0 = dt(u) = \sqrt{g^{00}(x)} \, n_\mu(x)\,  u^\mu\,,
\ee
where $dt$ denotes the 1-form obtained by differentiating the $t$ function. As another consequence of \eqref{u0nu}, we can re-write the right hand side of \eqref{hbd} in adapted coordinates as
\begin{equation}\label{hbd2}
  j^{0 \ldots 0, \mu_k, 0\ldots 0} (Q(t)) 
  \prod_{i\neq k} \frac{1}{\sqrt{g^{00}(Q_i(t))}}\,.
\end{equation}
If we parameterize the world lines by the time coordinate then $dQ_k^0/dt=1$, so we can re-write \eqref{hbd} as
\be\label{hbd3}
\frac{dQ_k^\mu}{dt} = 
\frac{j^{0 \ldots 0, \mu, 0\ldots 0}(Q(t))}
{j^{0 \ldots 0}(Q(t))}
\ee
with the index $\mu$ in the $k$-th place.

We also note the formula
\be\label{d4d3}
d_4(x) = \frac{1}{\sqrt{g^{00}(x)}}\, d_3(x)\,,
\ee
which follows from the fact that $d_4(x)$, the Lorentzian 4-volume spanned by the coordinate basis of $T_x\st$, equals the 4-volume spanned by the future-pointing vector $w\in T_x \st$ that is normal (in the sense of $g_{\mu\nu}$) to $\Sigma_t$ and has $w^0=1$, together with the 3 spacelike coordinate basis vectors. Due to orthogonality, this 4-volume is the product of the Lorentzian length of $w$ and the Riemannian 3-volume spanned by the 3 spacelike basis vectors, which is $d_3(x)$. That is, $d_4(x) = \sqrt{w^\mu w^\nu g_{\mu\nu}} \, d_3(x)$, while $w^\mu = (\nabla^\nu t \nabla_\nu t)^{-1} \nabla^\mu t$, which implies \eqref{d4d3}.

As a consequence of \eqref{alpha0g00} and \eqref{d4d3},
\be\label{d4alpha0d3}
d_4(x)\, \alpha^0(x) = d_3(x) \, I\,.
\ee

Now we are ready to determine the probability and current density. While the tensor $j^{\mu_1\ldots \mu_N}$ refers to invariant volume, we prefer to express all densities relative to coordinate volume because usually invariant volume, but not coordinate volume, becomes singular at a space-time singularity. We thus obtain from \eqref{denspsi0} the formula \eqref{denspsi} for the probability density. The corresponding formula \eqref{jpsi} for the current can be derived as follows. Regarding \eqref{denspsi} as known, we have $j_k^0(q)$ because it must equal the density $\dens(q)$; to see that this agrees with \eqref{jpsi}, use \eqref{d4alpha0d3}. The spacelike components can be obtained from the fact that the quotient of the current and the density is the velocity, and thus must be the same as in \eqref{hbd3}:
\be
\frac{j_k^\mu(Q)}{\dens(Q)} = \frac{dQ_k^\mu}{dt} = 
\frac{j^{0 \ldots 0, \mu, 0\ldots 0}(Q)}
{j^{0 \ldots 0}(Q)}\,.
\ee
As a consequence of \eqref{denspsi} and \eqref{d4alpha0d3},
\be
\dens(q) = d_4(q) \, j^{0\ldots 0}(q)
\ee
with
\be\label{d4def2}
d_4(q) = \prod_{k=1}^{\#q} \sqrt{-\det g_{\mu\nu}(q_k)}\,.
\ee
Hence, 
\be
j_k^\mu(q) = d_4(q)\, j^{0\ldots 0, \mu, 0 \ldots 0} 
\ee
with $\mu$ in the $k$-th place, which implies \eqref{jpsi}.


\section{Schwarzschild Space-Time}
\label{sec:schwarz}

As an example space-time $\st$, we use the Schwarzschild space-time, which we take to be Kruskal's maximal extension of the Schwarzschild metric \cite{Sch,Kru60,HE73,MTW}, and which features two spacelike singularities, one in the past and one in the future. 

\subsection{Definition}

We use the Kruskal coordinates $t',x',\vartheta,\varphi$, in which the metric is given by
\be\label{kruskalmetric}
ds^2 = F^2 \, dt^{\prime 2} - F^2\,dx^{\prime 2} - r^2(d\vartheta^2 + \sin^2 \vartheta \, d\varphi^2) 
\ee
with
\be
F^2 = \frac{16M^2}{r}e^{-r/2M}
\ee
and $r$ determined implicitly by the equation
\be\label{rdef}
t^{\prime 2} - x^{\prime 2} = -(r-2M) e^{r/2M}\,.
\ee
The coordinates $t'$ and $x'$ only take such values that
\be
t^{\prime 2} - x^{\prime 2} \leq 2M\,.
\ee
We note for later use that
\be\label{dschwarz}
d_4(t',x',\vartheta,\varphi) = F^2r^2\sin\vartheta\,, \quad
d_3(t',x',\vartheta,\varphi) = Fr^2\sin\vartheta\,.
\ee

The singularity lies at $t^{\prime 2} - x^{\prime 2} = 2M$. That is, the space-time $\st$ is the manifold-with-boundary given by
\be
\st = \bigl\{(t',x')\in \RRR^2: t^{\prime 2} - x^{\prime 2} \leq 2M\bigr\} \times \SSS^2\,,
\ee
and the singularity $\sing=\partial \st$ is
\be
\sing = \bigl\{(t',x'): t^{\prime 2} - x^{\prime 2} = 2M\bigr\} \times \SSS^2\,,
\ee
which has two connected components, $\sing=\sing_1 \cup \sing_2$, with
\be
\sing_1 = \{(t',x'):t' = \sqrt{2M+x^{\prime 2}} \} \times \SSS^2
\ee
and
\be
\sing_2 = \{(t',x'):t' = -\sqrt{2M+x^{\prime 2}} \} \times \SSS^2\,.
\ee
$\sing_1$ is a future singularity, and $\sing_2$ a past singularity.

A curvature cut-off can be implemented by cutting out from $\st$ a neighborhood of the singularity, thus making a spacelike hypersurface $\surface$ the new boundary. For example, in order to cut out $\sing_1$ from the Schwarzschild space-time we could set
\be\label{cutoffschwarz}
\surface = \{(t',x'):t' = \sqrt{2M+x^{\prime 2}}-\varepsilon\}\times \SSS^2
\ee
with $\varepsilon>0$ a small constant.


As the time foliation $\foliation$ we take the level surfaces of the $t'$ function. Note that these hypersurfaces are not Cauchy surfaces, except for $-\sqrt{2M}<t'<\sqrt{2M}$. For $t'\geq \sqrt{2M}$, $\Sigma_{t'}$ \emph{borders on the singularity} $\sing_1$, as $\sing_1\cap\Sigma_{t'}\neq \emptyset$, and $\Sigma_{t'}$ consists of two connected components,\footnote{Needless to say, particles in different components can be entangled with each other. As a consequence, the Bohmian velocity of one particle may depend on the position of the other, and results of experiments carried out in different components can be nonlocally correlated.} corresponding to $x'\geq\sqrt{t^{\prime 2}-2M}$ and $x'\leq-\sqrt{t^{\prime 2}-2M}$, each of topology $[0,\infty)\times \SSS^2$. For $t'>\sqrt{2M}$, the singular boundary $\partial \Sigma_{t'}$ consists of two unconnected spheres, one corresponding to $x'=\sqrt{t^{\prime 2}-2M}$, the other to $x'=-\sqrt{t^{\prime 2}-2M}$; for $t'=\sqrt{2M}$ the two spheres coincide. Likewise, for $t' \leq -\sqrt{2M}$, $\Sigma_{t'}$ borders on the singularity $\sing_2$ and consists of two connected components, each of topology $\RRR^3$ minus an open ball; for $t'<-\sqrt{2M}$, the singular boundary $\partial \Sigma_{t'}$ consists of two unconnected spheres.

Among other spacelike foliations, there are some whose leaves will, like those of $\foliation$, border on the singularity and others whose leaves will not, e.g., $\foliation = \{\Sigma_s: -1<s<1\}$ with
\be\label{foliation2}
\Sigma_s = \{  t' = s\sqrt{2M+x^{\prime 2}} \} \,.
\ee
Indeed, all of these leaves are Cauchy surfaces; as a consequence, such foliations will be uninteresting to us, as the time evolution is unitary and the pure-to-mixed evolution we introduced does not occur.

\subsection{End Points of Causal Curves}
\label{sec:endpoints}

Finally, we mention that every causal curve (i.e., one that is everywhere timelike or lightlike), if it is future inextendible and does not reach infinity, has an end point on the singularity $\sing_1$. In particular, it is not possible that the curve has more than one accumulation point on $\sing_1$. 

To see this, consider such a curve $x(t')$ that cannot be extended beyond time $t_0'$; to see that the $x'$ coordinate converges as $t'\to t'_0$, note that for any $\varepsilon>0$ and $t_0'-\varepsilon =:t'_1 < t' < t'_0$, $x'(t')$ must lie between $x'(t'_1)-\varepsilon$ and $x'(t'_1)+\varepsilon$. For showing that also the angular coordinates converge it suffices, by rotational symmetry, to consider the $\varphi$ coordinate and show that the function $\varphi(t')$ has bounded variation in the open time interval $(t'_1,t'_0)$. The total variation of $\varphi(t')$ in this interval is
\be\label{variation}
V=\int_{t'_1}^{t'_0} dt' \, \Bigl|\frac{d\varphi}{dt'}\Bigr|\,.
\ee
Since the curve is causal, $ds^2\geq 0$, we have from \eqref{kruskalmetric} that
\be\label{est1}
\Bigl|\frac{d\varphi}{dt'}\Bigr| \leq \frac{F}{r}=\frac{4M}{r^{3/2}}e^{-r/4M}\,.
\ee
We find it useful to parameterize the curve by $r$ rather than $t'$; to this end, we obtain from \eqref{rdef} (noting that $dr<0$) that 
\be\label{est2}
dt' = -\frac{re^{r/2M}}{2M(2t'+2x' \,dx'/dt')}dr \leq 
-\frac{re^{r/2M}}{4M (t'_1-|x'(t'_1)|-\varepsilon)}dr
\ee
using $|dx'/dt'|\leq 1$ and choosing $\varepsilon$ so small that $t'+x' \,dx'/dt'\geq t'_1-|x'(t'_1)|-\varepsilon>0$. Inserting \eqref{est1} and \eqref{est2} into \eqref{variation} and using $r\geq 0$, we obtain that
\begin{align}
V\leq& \int_0^{r(t'_1)} dr\, \frac{4M}{r^{3/2}}e^{-r/4M} 
\frac{re^{r/2M}}{4M(t'_1-|x'(t'_1)|-\varepsilon)} \nonumber\\
\leq& \frac{e^{r(t'_1)/4M}}{(t'_1-|x'(t'_1)|-\varepsilon)}
\int_0^{r(t'_1)} \frac{dr}{\sqrt{r}} 
= \frac{e^{r(t'_1)/4M}}{(t'_1-|x'(t'_1)|-\varepsilon)} 2\sqrt{r(t'_1)} <\infty \,,
\end{align}
which completes the proof.

\section{Future Spacelike Singularities}
\label{sec:future}

\subsection{Mathematical Framework}
\label{sec:framework}

We now set up the mathematical structure; since we do not strive for mathematical precision, not every concept will be sharply defined. 

We assume that we are given a 4-manifold with boundary $\st$ as the space-time (where $\sing=\partial \st$ is the boundary), equipped with a Lorentzian metric $g_{\mu\nu}$ on the interior $\st^\circ=\st\setminus \sing$, a time orientation, a complex vector bundle over $\st$ of spin spaces which we denote by $\spin = \cup_{x\in\st}\spin_x$, and a foliation into spacelike hypersurfaces, the time foliation $\foliation$. We assume that $\sing=\partial \st$ is a future singularity. For simplicity, we assume further that $\foliation$ is parameterized by a time parameter $t$, i.e., $\foliation = \{\Sigma_t: t_1<t<t_2\}$, and that this is done in such a way that $t$ as a function on $\st$ (defined by $t(x)=\tau \Leftrightarrow x\in\Sigma_\tau$) has nowhere-vanishing gradient, $\forall x\in\st:\: \nabla t(x) \neq 0$. For the Dirac equation, $\spin$ has complex rank 4 over $\st$. 

Let
\be
\conf=\conf(\Sigma) = \bigcup_{n=0}^\infty \Sigma^n
\ee
be the configuration space of a variable number of particles associated with the hypersurface $\Sigma$; the union is understood as a disjoint union; elements of $\conf$ are ordered configurations, but the ordering is physically irrelevant. We write $\#q$ for the number of particles in the configuration $q$. On the set $\conf$ we consider two measures, the Riemannian (invariant) volume $d_3(q)\,dq$ and the coordinate volume $dq$, with $d_3$ as in \eqref{d3def2}. They arise from two measures on $\Sigma$, the Riemannian (invariant) volume $d_3(x)\,dx$ and the coordinate volume $dx$, with $d_3(x) = \sqrt{-\det g^{(3)}(x)}$. The measure $d_3(x)\,dx$ is the one that arises from the Riemannian metric $g^{(3)}$ on $\Sigma$ inherited from $g_{\mu\nu}$.

The manifold $\conf$ is further equipped with the vector bundle $\cup_{q\in\conf} \spin_q$ defined in \eqref{spinqdef} from the bundle $\cup_{x\in\Sigma}\spin_x$. Throughout this paper the probability density $\dens$ on $\conf$ and the probability current $j$ on $\conf$ are \emph{permutation-invariant}, i.e., $\dens(x_{\sigma 1}, \ldots, x_{\sigma n})=\dens(x_1,\ldots,x_n)$ and $j_{\sigma k}(x_{\sigma 1}, \ldots, x_{\sigma n})=j_k(x_1,\ldots,x_n)$ for any permutation $\sigma$ of $\{1,\ldots,n\}$, with $j_k\in T_{x_k}\Sigma$ the component of $j$ associated with the $k$-th particle.

We now define the relevant Hilbert space. The 1-particle Hilbert space associated with the time leaf $\Sigma$ consists of square-integrable cross-sections of the spin bundle, $\psi:\Sigma\to \cup_{x\in\Sigma}\spin_x$, relative to the measure $d_3(x)\, dx$. For this space we write
\be
\Hilbert_1 = L^2\bigl(\Sigma,\cup_{x\in\Sigma}\spin_x,d_3(x)\,dx\bigr)\,.
\ee
For any 1-particle Hilbert space $\Hilbert_1$, let $\Gamma_\fer(\Hilbert_1)$ and $\Gamma_\bos(\Hilbert_1)$ denote the fermionic and bosonic Fock space over $\Hilbert_1$, respectively; i.e.,
\be
\Gamma_\fer(\Hilbert_1) = \bigoplus_{n=0}^\infty S_- \Hilbert_1^{\otimes n}\,, \quad
\Gamma_\bos(\Hilbert_1) = \bigoplus_{n=0}^\infty S_+ \Hilbert_1^{\otimes n}\,,
\ee
where $S_-$ and $S_+$ are the anti-symmetrizer and the symmetrizer, respectively. We write $\Gamma_\eith$ for either $\Gamma_\fer$ or $\Gamma_\bos$.
The relevant Hilbert spaces for us will be
\be
\Hilbert_\Sigma = \Gamma_\eith
\bigl(L^2(\Sigma,\cup_{x\in\Sigma}\spin_x,d_3(x)\, dx)\bigr)
\ee
for $\Sigma\in\foliation$. A vector $\psi\in\Hilbert_\Sigma$ can be regarded as a cross-section of the bundle $\cup_q\spin_q$, i.e., $\psi: \conf \to\cup_{q\in\conf} \spin_q$ with $\psi(q)\in \spin_q$. In these terms, the inner product of $\Hilbert_\Sigma$ can be expressed as
\be
\scp{\phi}{\psi} = \int_{\conf} dq\, d_3(q) \, \phi^*(q) \, \psi(q)\,.
\ee

Every density matrix $\dm$ on $\Hilbert_\Sigma$ can be expressed as a function $\dm(q;\aq)$ of two ordered configuration variables. In fact, $\dm(\cdot;\cdot)$ is a cross-section of the bundle $\cup_{q,\aq} Hom(\spin_{\aq}, \spin_q)$ over the base manifold $\conf\times \conf$, where $Hom(\spin_{\aq}, \spin_q)$ denotes the space of $\CCC$-linear mappings $\spin_{\aq}\to\spin_q$.

\bigskip

We are now ready to turn again to the quasi-Lindblad form as in \eqref{qLin}, and to point out in which way \eqref{dmevol} is of this form, provided we can leave aside the complications arising from the limit $y\to x,y\notin\sing$ and evaluate the density matrix directly on the singularity. Set $\lambda=(s,x)$ with $s\in \{1,2,3,4\}$ a spin index and $x\in \sing_t$ a point on the singularity; use the measure $\mu_t=\# \otimes d^2x\, w(x)$, where $\#$ denotes the counting measure on $\{1,2,3,4\}$; that is,
\be
\int \mu_t(d\lambda) f(\lambda) = \sum_{s=1}^4 \int_{\sing_t} d^2x \, w(x) \, f(s,x)\,.
\ee
Let $\{b_1(x),\ldots,b_4(x)\}$ be an orthonormal basis of $\spin_x$ consisting of eigenvectors of the positive definite Hermitian endomorphism
\be\label{alphasingdef}
\alpha_\sing(x) = \lim_{y \to x,y \notin\sing} c_\mu(x) \, d_4(y) \alpha^\mu(y)
\ee
of $\spin_x$ with eigenvalues $a_1(x),\ldots,a_4(x)$. Finally, let $\hat A_t(s,x)$ be $\sqrt{a_s(x)}$ times the annihilation operator on Fock space that annihilates a particle with position $x$ and spin $b_s(x)$:
\be\label{LinA}
\bigl(\hat A_t(s,x) \psi \bigr)(q) = \sqrt{a_s(x)}\sqrt{\#q +1} \, b_s^*(x) \, \psi(q,x)\,.
\ee
Then
\be
\bigl(\hat A_t(s,x) \dm \hat A^*_t(s,x)\bigr)(q;\aq) = 
\sqrt{(\#q+1)(\#\aq+1)} \,  b_s^*(x) \, \dm(q,x;\aq,x) \,a_s(x) \, b_s(x)\,,
\ee
and thus
\be
\sum_{s=1}^4\bigl(\hat A_t(s,x) \dm \hat A^*_t(s,x)\bigr)(q;\aq) = 
\sqrt{(\#q+1)(\#\aq+1)}  \, 
\tr_{\spin_x} \bigl( \dm(q,x;\aq,x) \, \alpha_\sing(x)\bigr)\,,
\ee
which brings \eqref{dmevol} into the form \eqref{qLin}.

\subsection{Derivation From Equivariance}

We now give a derivation of \eqref{Lop}, and thus of the explicit evolution equation of the density matrix, from the framework of Bohmian mechanics, in particular from the requirements of equivariance and independence of disentangled systems, and from the assumption that the evolution of the density matrix is linear (i.e., of the form \eqref{dmevol} with unspecified $\Lop$). 

\begin{enumerate}
\item We assume that particles move according to
\be\label{Bohm3}
\frac{dQ_i}{dt} = v_i(t,Q(t))\,,
\ee
where $v(t,q)$ is a time-dependent permutation-invariant vector field on configuration space $\conf(\Sigma_t)$, and $i$ runs through the $3\#Q$ dimensions of $\conf(\Sigma_t)$ at $Q$. Here we use a coordinate system whose time component is given by the parameter $t$ of the time foliation and that is otherwise arbitrary. (The vector field $v$ depends on the choice of coordinates.) 

The motion according to \eqref{Bohm3} continues until the configuration hits the boundary of configuration space, which means that one (or more) of the particles hits (or hit) the singularity; in this event that particle gets (or those particles get) annihilated, i.e., removed from the configuration, according to \eqref{Qjump}.

\item We assume that $\sing_t=\Sigma_t\cap \sing$ is 2-dimensional and can be regarded as the boundary of $\Sigma_t$. (In our Schwarzschild example,
\be
\sing_{t'} = \Sigma_{t'}\cap S= 
\bigl\{t'=\mathrm{const.}, x'=\pm\sqrt{t^{\prime 2}-2M}\bigr\}
\ee
when $t'>\sqrt{2M}$; this is a disjoint union of two 2-spheres, and is indeed the boundary of $\Sigma_{t'}$, a disjoint union of two items of topology $[0,\infty)\times \SSS^2$.) 

Let $v_\sing (t,x)$ denote the speed at which the singular boundary $\sing_t$, regarded as a time-dependent 2-surface in coordinate 3-space, moves with increasing $t$ in the surface-normal direction at the coordinate point representing $x\in \sing_t$; since $\sing$ is spacelike, this speed must be greater than the speed of light. This speed can be expressed in terms of the function $T(\xi)$ that specifies the time coordinate at which a certain point $\xi$ in coordinate 3-space is reached by the singularity. (In our Schwarzschild example, if $\xi=(x',\vartheta,\varphi)$ then $T(\xi)=\sqrt{x^{\prime 2}+2M}$.) Indeed, 
\be
v_\sing (T(\xi),\xi) = \frac{1}{|\grad  T(\xi)|}\,,
\ee
where $|\cdot|$ is the Euclidean norm in coordinate 3-space and $\grad  T$ the gradient of $T$ in coordinate 3-space. To see this, note that a line in coordinate space-time starting at $(T(\xi),\xi)$ on the singularity with direction $(|\grad T|,\grad T/|\grad T|)$ will be tangent to the singularity. Note also that $\grad T(\xi)$ is orthogonal, in coordinate 3-space, to $\sing_{T(\xi)}$. (In our Schwarzschild example, $\grad  T = x'/\sqrt{x^{\prime 2}+2M}\,  \partial_{x'}$, pointing in the radial direction, and $v_\sing (T(\xi),\xi)= \sqrt{x^{\prime 2}+2M}/|x'|$.)

It now follows that the continuity equation for a permutation-invariant distribution density $\dens$ (relative to coordinate volume) of the configuration reads
\begin{align}
\frac{\partial \dens}{\partial t}(t,q) =& \nonumber 
-\sum_{i=1}^{3\#q} \partial_i \bigl( \dens(t,q) \,v_i(t,q) \bigr) \:+\\ 
&+(\#q+1) \int_{\sing_t} d^2x \, \dens(t,q, x) \, 
\bigl(v_\sing (t,x)-v_{x,\perp}(t,q, x)\bigr)\,,
\label{conti3}
\end{align}
where $d^2x$ is the surface area element in coordinate space (in our Schwarzschild example, $d^2x= d\vartheta \, d\varphi$), and $v_{x,\perp}$ denotes the component of the velocity of the particle at $x$ that is orthogonal, in coordinates, to $\sing_t$ and inward-pointing (i.e., away from the singularity). It can be expressed as
\be
v_{x,\perp} = \frac{\grad T}{|\grad T|} \cdot (v_{x1},v_{x2},v_{x3})\,.
\ee

The first term on the right hand side of \eqref{conti3}, the negative spatial divergence of a probability current, represents the change in density due to the flow with velocity $v(t,q)$, and the second term represents the gain in density due to jumps in configuration space to the configuration $q$ from configurations containing one further particle at location $x$, where $x$ lies on the singularity. To understand what the second term must be, keep $q$ fixed, note that there are $\#q+1$ possibilities for the position of the variable $x$ among the variables of $q$, consider configurations of the form $(q, x)$ with $x$ near the singularity, and note that the particle at $x$ will be swallowed by the singularity within the next $dt$ seconds if and only if its distance from the singularity, in coordinates, is less than $\bigl(v_\sing (t,x)-v_{x,\perp}(t,q, x)\bigr)dt$. Thus, the second term is the amount of $\dens(t,q, x)$, for fixed $q$ and arbitrary $x$, that flows into the singularity within the next $dt$ seconds.

\item We assume that the velocity vector field $v(t,q)$ is given, as usual, as the quotient of a current vector field and a density function,
\be\label{vjdens}
v(t,q) = \frac{j(t,q)}{\dens(t,q)}\,.
\ee

\item We anticipate that the current $j$ and the density $\dens$ are determined not by a wave function but by a density matrix $\dm_t$, $j(t,q)=j^{\dm_t}(q)$ and $\dens(t,q) = \dens^{\dm_t}(q)$. Moreover, we assume that they are given in terms of $\dm_t$ by the formulas \eqref{densdm1} and \eqref{jdm1}. These formulas are the obvious extensions from wave functions to density matrices of the formulas \eqref{denspsi} and \eqref{jpsi} usually utilized in Bohmian mechanics. In particular, for a pure state $\dm=\pr{\psi}$, \eqref{densdm1} and \eqref{jdm1} reduce to \eqref{denspsi} and \eqref{jpsi}.

\item We assume that the density matrix $\dm_t$ obeys, as in quantum dynamical semigroups, a linear evolution, which we write
\be\label{dmevol2}
\frac{\partial \dm_t}{\partial t} = \tfrac{i}{\hbar} [\dm_t,\hat H_t] + \Lop \dm_t\,.
\ee
This is literally the same equation as \eqref{dmevol}, but so far with $\Lop$ unspecified. The operator $\hat H_t$ is the Dirac Hamiltonian; we have separated this term knowing that it remains in the absence of the singularity $\sing$, and thus know that $\Lop=0$ in the absence of a singularity. As a consequence of \eqref{dmevol}, $\dens^{\dm_t}$ evolves as follows:
\be\label{conti4}
\frac{\partial \dens^{\dm_t}}{\partial t} (q) =
-\sum_{i=1}^{3\#q} \partial_i j_i^{\dm_t} 
+ d_3(q) \, 
\tr_{\spin_q}\bigl((\Lop \dm_t) (q;q) \bigr) \,.
\ee
Here we have used the known continuity equation for the Dirac equation in a curved space-time, a key element of the proof of equivariance of Bohmian mechanics in a curved space-time \cite{3forms,Tum06d}.

\item We want equivariance, i.e., $\dens_t=\dens^{\dm_t}$ for all $t$ if initially. For this, we need that the right hand sides of equations \eqref{conti3} and \eqref{conti4} coincide when $\dens_t=\dens^{\dm_t}$ is assumed; since the first terms (the divergence of the current) do coincide in that case, we only need that
\begin{multline}\label{Lop5}
\tr_{\spin_q}(\Lop \dm_t) (q;q)=(\#q+1)  \:\times\\
\times \int_{\sing_t} d^2x \, \lim_{y\to x, y\notin \sing} 
\tr_{\spin_{q,y}}\Bigl(\dm_t(q,y;q,y) \bigl(d_3(y)\, v_\sing (t,x)  \, I
- d_4(y) \alpha^\perp(y)\bigr) \Bigr)\,,
\end{multline}
where
\be
\alpha^\perp(y)= 
\frac{\grad T}{|\grad T|}\cdot \bigl(\alpha^1(y), \alpha^2(y),\alpha^3(y)\bigr)
\ee
is the component of $\alpha^\mu(y)$ that is tangent to $\Sigma_t$ and orthogonal, in coordinates, to $\sing_t$ at $x$.

If \eqref{Lop5} is the case, we can argue that as soon as $\dens_t=\dens^{\dm_t}$ for one $t$, then $\partial \dens_t/\partial t = \partial \dens^{\dm_t}/\partial t$, and thus $\dens_{t+dt} = \dens^{\dm_{t+dt}}$, and so on into the future. Put more mathematically, we take for granted that the continuity equation \eqref{conti3}, as a PDE for $\dens_t(q)$, has unique solutions for every initial condition, observe that $(t,q)\mapsto \dens^{\dm_t}(q)$ is a solution of \eqref{conti3} by virtue of \eqref{conti4} and \eqref{Lop5}, and conclude that if $\dens$ agrees with $\dens^{\dm_t}$ for one $t$ then it must also agree at every later time.

\item The obvious choice of $\Lop$ that will make \eqref{Lop5} true is
\begin{multline}
(\Lop \dm_t) (q;\aq) =
(\#q+1)^{1/2}(\#\aq+1)^{1/2} \:\times\\
\times \int_{\sing_t} d^2x \, \lim_{y\to x, y\notin \sing}
\tr_{\spin_{y}}\Bigl(\dm_t(q,y;\aq,y) \bigl(d_3(y)\, v_\sing (t,x)  \, I
- d_4(y)\,\alpha^\perp(y)\bigr) \Bigr)\,. \label{Lop6}
\end{multline}
The only differences between \eqref{Lop5} and \eqref{Lop6} are the trace over $\spin_q$ and that \eqref{Lop5} has $q$ inserted for $\aq$.

Indeed, \eqref{Lop6} is strongly suggested by the wish that disentangled systems should behave independently: Consider two disentangled systems, the first consisting of just one particle that is about to hit the singularity, the second consisting of several particles far away from the singularity. Then
\be
\dm_t(q,y;\aq,y) = \dm_t^{(1)}(y;y) \otimes \dm_t^{(2)}(q;\aq)\,,
\ee
with the superscript indicating the system, and the contribution to $\dm_t$ with one particle less, $\Lop\dm_t(q;\aq)$, arising from the particle of the first system hitting the singularity, should be proportional to $\dm_t^{(2)}(q;\aq)$. This is the case according to \eqref{Lop6}.

\item Eq.~\eqref{Lop6} agrees with \eqref{Lop}: Note that, in coordinates,
\be
c_\mu(x) = \Bigl(\frac{1}{\sqrt{1+\kappa^2}},
\frac{-\grad  T}{\sqrt{1+\kappa^2}} \Bigr)
\ee
with $\kappa=|\grad  T|$;
\begin{align}
w(x) &= \sqrt{1+\kappa^{-2}} \,;\\
w(x) \,c_\mu(x)\,  d_4(y)\,\alpha^\mu(y) &= \nonumber 
\sqrt{\frac{1+\kappa^{-2}}{1+\kappa^2}} \frac{\kappa}{\kappa} 
\Bigl(d_4(y)\,\alpha^0(y) - d_4(y)\,
\grad \, T \cdot (\alpha^1,\alpha^2,\alpha^3)\Bigr)\\
&= \nonumber 
\sqrt{\frac{\kappa^2+1}{1+\kappa^2}} \Bigl(\frac{d_3(y)}{\kappa}\,I -
d_4(y) \, \frac{\grad \, T}{\kappa} \cdot (\alpha^1,\alpha^2,\alpha^3) \Bigr)\\
&= 
d_3(y)\, v_\sing(t,x)\,I -
d_4(y) \, \alpha^\perp(y)\,.
\end{align}
\end{enumerate}

This concludes our derivation. We have shown in particular that the distribution $\dens^{\dm_t}$ as in \eqref{densdm1} is equivariant with respect to the evolution of the Bohmian particles.

\bigskip

In Schwarzschild space-time for $t'\geq \sqrt{2M}$ the evolution equation \eqref{dmevol} reads explicitly:
\begin{align}\nonumber
\frac{\partial \dm_{t'}(q;\aq)}{\partial t'} =&
-\tfrac{i}{\hbar} \hat H_{t'}(q) \dm_{t'}(q;\aq)
+\tfrac{i}{\hbar} \hat H_{t'}(\aq) \dm_{t'}(q;\aq)\\
&+4M\sqrt{(\#q+1)(\#\aq+1)} \sum_{\sigma=\pm 1} 
\int_0^\pi d\vartheta\, \sin\vartheta\int_0^{2\pi} d\varphi \, 
\lim_{y'\searrow \sqrt{t^{\prime 2}-2M}} r^{3/2}(t', y') \:\times \nonumber\\
&\times\: \tr_{\spin_{y'}}\Biggl(\dm_{t'}\bigl(q,(\sigma y',\vartheta,\varphi);\aq,(\sigma y',\vartheta,\varphi)\bigr) \Bigl(\frac{t'}{\sqrt{t^{\prime 2}-2M}}  \, I
- F(t',y')\,\alpha^1(t',\sigma y')\Bigr) \Biggr)
\label{dmschwarz}
\end{align}
with $\spin_{y'}$ the spin space at $(t',y',\vartheta,\varphi)$ and
\be
F(t',y')\,\alpha^1(t',\sigma y') =  
\begin{pmatrix}
0&0&0&1\\
0&0&1&0\\
0&1&0&0\\
1&0&0&0
\end{pmatrix}
\ee
relative to the orthonormal basis of $\spin_{y'}$ associated with the orthonormal basis
\be
(F^{-1}\partial_{t'},F^{-1}\partial_{x'},r^{-1}\partial_{\vartheta},
(r\sin\vartheta)^{-1}\partial_{\varphi})
\ee
of $T_{y'}\st$.
Note that all terms in \eqref{dmschwarz} of the form $\exp(r(t',y')/2M)$ could be dropped since $r\to 0$ as $y'\to \sqrt{t^{\prime 2}-2M}$. Another remark concerns the term $r^{3/2}$ in \eqref{dmschwarz}: This factor, which is essentially $d_3(y)$, tends to zero as $y$ approaches the singularity; since the density in coordinates is conserved and thus cannot tend to zero when reaching the singularity, we must conclude that the density relative to the invariant measure, as given by $\tr_{\spin_y} \dm_t(q,y;\aq,y)$, diverges at the singularity at the rate $r^{-3/2}$.

\bigskip

We close this subsection with a remark about the impossibility of a unitary evolution in the presence of a spacelike singularity. The process of the particle configuration has continuity equation \eqref{conti3}, while a unitary evolution of a wave function,
\be
i\hbar \frac{\partial \psi}{\partial t} = \hat H_0 \psi + \hat H_1 \psi\,,
\ee
where $\hat H_0$ is the free Dirac Hamiltonian and $\hat H_1$ a putative further term, would imply for $\dens^{\psi_t}(q) =d_3(t,q)\, \psi_t^*(q) \, \psi_t(q)$ that
\be\label{conti2}
\frac{\partial \dens^{\psi_t}}{\partial t} = -\sum_{i=1}^{3\#q} \partial_i j_i^{\psi_t} +
\tfrac{2d_3(t,q)}{\hbar} \Im \, \psi_t^*(q) (\hat H_1 \psi_t)(q)\,.
\ee 
For equivariance we need this equation to coincide with \eqref{conti3}, and using \eqref{vjdens} we find that we need that
\begin{multline}
(\#q+1) \int_{\sing_t} d^2x \, d_3(x) \, \psi_t^*(q,x) \, \psi_t(q,x) \, 
\bigl(v_\sing (t,x)-v_{x,\perp}(t,q, x)\bigr)=\\
\tfrac{2}{\hbar} \Im \, \psi_t^*(q) (\hat H_1 \psi_t)(q)\,.
\end{multline}
But the last equation cannot hold, no matter how we choose $\hat H_1$, as the left hand side contains the factor $\psi_t^*(q,x)$ but the right hand side has $\psi_t^*(q)$ instead. In the case of timelike singularities, a similar problem can be solved by imposing a quasi-boundary condition on $\psi_t$ that will ensure a relation between $\psi_t(q)$ and $\psi_t(q,x)$; however, for a future spacelike singularity, there is no room for a boundary condition, since $\psi_t(q,x)$ is determined by the Dirac evolution.

\subsection{Quasi-Lindblad Equation for Reduced Density Matrix}
\label{sec:reduced}

The quasi-Lindblad equation \eqref{dmevol} with \eqref{Lop} also arises in another context, in which the singularity $\sing$ is replaced by a spacelike hypersurface $\surface$: then \eqref{dmevol} with \eqref{Lop} describes the time evolution of the reduced density matrix of what has not yet passed the hypersurface $\surface$. So consider a space-time without singularities with a time foliation $\{\Sigma_t\}$ and therein a spacelike hypersurface $\surface$ such that $\surface_t=\Sigma_t\cap \surface$ is always 2-dimensional.

As a concrete simple example, readers may think of Minkowski space-time
\be\label{exS}
(\st,g) = \bigl(\RRR^4,\diag(1,-1,-1,-1)\bigr)
\ee
and the time foliation defined by a Lorentz frame, so that $\Sigma_t$ are parallel spacelike hyperplanes (where the time coordinate assumes the value $t$). Let $\surface$ be another spacelike hyperplane, not parallel to the $\Sigma_t$. We first describe the general definitions and then illustrate them using this example situation.

Let $\psi: \cup_{\Sigma\in\foliation} \Sigma^N\to \cup_{\Sigma\in\foliation}\cup_{q\in\Sigma^N} \spin_q$ be a fermionic or bosonic $N$-particle (Dirac) wave function evolving, for example, without interaction. Let $J^+(\surface)$ denote the future of $\surface$ and $J^-(\surface)$ its past. For every $\Sigma \in \foliation$, set $\Sigma^\pm=\Sigma \cap J^\pm(\surface)$. Let $\foliation^\pm = \{\Sigma^\pm : \Sigma \in \foliation\}$, which is a foliation of $J^\pm(\surface)$. It is a basic fact about (fermionic or bosonic) Fock spaces $\Gamma_\eith$ that
\be
\Gamma_\eith(\Hilbert_{1a} \oplus \Hilbert_{1b}) = 
\Gamma_\eith(\Hilbert_{1a}) \otimes \Gamma_\eith(\Hilbert_{1b})\,.
\ee
As a consequence, for disjoint subsets $A,B$ of 3-space, $A\cap B = \emptyset$,
\be
\Gamma_\eith(L^2(A \cup B)) = \Gamma_\eith(L^2(A)) \otimes \Gamma_\eith(L^2(B))\,.
\ee
Since $\Sigma\setminus (\Sigma^+ \cup \Sigma^-) = \Sigma\cap \surface$ is a null set by assumption and thus not relevant to square-integrable functions, we have that, for every $\Sigma\in \foliation$,
\be
\Hilbert_\Sigma = \Hilbert_{\Sigma^+} \otimes \Hilbert_{\Sigma^-}\,.
\ee
Now define
\be\label{dm-def}
\dm^- = \tr_+ \pr{\psi_\Sigma}\,,
\ee
where $\tr_+$ means the partial trace over $\Hilbert_{\Sigma^+}$; $\dm^-$ is a density operator on $\Hilbert_{\Sigma^-}$. 

We note that, as a consequence of the fact that $\psi$ lies in the $N$-particle sector of Fock space, $\dm^-$ is block-diagonal relative to the particle number sectors of Fock space, i.e., $\scp{\phi}{\dm^-|\chi}=0$ whenever $\phi$ lies in the $n$-particle sector of $\Hilbert_{\Sigma^-}$ and $\chi$ in the $m$-particle sector with $m\neq n$. To see this, note that this is the case for $\pr{\psi_\Sigma}$ by assumption, and the partial trace can be carried out using an orthonormal basis $\{b_i\}$ of $\Hilbert_{\Sigma^+}$ that consists of basis vectors that are eigenvectors of particle number, so that $\scp{\phi}{\dm^-|\chi} = \sum_i \scp{\phi\otimes b_i}{\psi_\Sigma} \scp{\psi_\Sigma}{\chi \otimes b_i}$.

We can obtain a more explicit expression for $\dm^-$ by writing it as a function $\dm^-(q;\aq)$; namely, with $0\leq n \leq N$ and $q,\aq \in (\Sigma^-)^n$,
\be\label{dmint}
\dm^-(q;\aq) = 
\binom{N}{n}
\int\limits_{(\Sigma^+)^{N-n}} d\cq\, d_3(\cq) \:
\tr_{\spin_{\cq}}\bigl(\psi(q,\cq) \, \psi^*(\aq,\cq)\bigr)
\ee
with $d_3(\cq)$ the 3-volume factor as in \eqref{d3def2}, $\psi=\psi_\Sigma$,
and $\tr_{\spin_{\cq}}$ the partial trace over those spin indices belonging to particles in $\cq$. Here we use that
\be
\spin_{q, \cq} = \spin_{q} \otimes \spin_{\cq}\,.
\ee
The binomial factor in \eqref{dmint} arises from the re-ordering of variables in $\psi$ so that the $N-n$ variables of $\cq$ appear last. 
In case $n=N$, we take $(\Sigma^+)^{N-n}=(\Sigma^+)^0$ to be a one-element set and the 0-dimensional integration measure to give measure 1 to that one element, so that the integral equals $\psi(q) \, \psi^*(\aq)$. In other words, the block $\dm_N^-$ of $\dm^-$ in the $N$-particle sector of $\Hilbert_{\Sigma^-}$ is just $\hat P \pr{\psi} \hat P$ with $\hat P:L^2(\Sigma^N) \to L^2((\Sigma^-)^N)$ the projection to the subspace in which all particles lie in $\Sigma^-$. 

Let us formulate the time evolution of $\dm^-$. For an expression like $\partial\dm_t^-(q;\aq)/\partial t$ to make sense, we regard now $q$ and $\aq$ not as points on $\Sigma_t$ but as their spatial coordinates; for simplicity, we will write $\Sigma_t$ for the image of $\Sigma_t$ in coordinate 3-space. We need to differentiate \eqref{dmint} with respect to time, and thus to differentiate an integral with time-dependent domain, as in
\be
g(t,x) = \int_{B_t} dy\, f(t,x,y)\,,
\ee
where $x$ is a variable in $\RRR^n$, $y$ a variable in $\RRR^m$, and $B_t\subseteq \RRR^m$ a set with smooth boundary $\partial B_t$ moving in a smooth way. The rule we need can be regarded as a version of the fundamental theorem of calculus, which in its simplest form reads
\be\label{fundthm1}
\frac{d}{dt} \int_0^t du\,f(u)=f(t)
\ee
while the form we need reads
\be\label{fundthm2}
\frac{\partial g}{\partial t} = \int_{B_t} dy\,\frac{\partial f}{\partial t}+ 
\int_{\partial B_t} dy\, v_{B}(t,y) \, f(t,x,y)\,,
\ee
where $v_B(t,y)$ is the (signed) speed at which the surface $\partial B_t$ moves outward in the direction orthogonal to the surface at $y$.
We thus obtain, with $B_t = (\Sigma_t^+)^{N-n}$,
\begin{align}
\frac{\partial\dm_t^-(q;\aq)}{\partial t}
&=\binom{N}{n}\int\limits_{(\Sigma_t^+)^{N-n}} d\cq\, 
\frac{\partial d_3(t,\cq)}{\partial t}\, \tr_{\spin_{\cq}} \,\psi_t(q,\cq) \, 
\psi_t^*(\aq,\cq) \:+\nonumber\\
&+\binom{N}{n}\int\limits_{(\Sigma_t^+)^{N-n}} d\cq \, d_3(\cq)\,
\tr_{\spin_{\cq}}(-\tfrac{i}{\hbar}\hat H_t\psi_t)(q,\cq) \, 
\psi_t^*(\aq,\cq) \:+\nonumber\\
&+\binom{N}{n}\int\limits_{(\Sigma_t^+)^{N-n}} d\cq \, d_3(\cq)\,
\tr_{\spin_{\cq}} \,\psi_t(q,\cq) \, 
\tfrac{i}{\hbar} (\hat H_t \psi_t)^*(\aq,\cq) \:+ \nonumber\\
&+ \binom{N}{n}\int\limits_{\partial (\Sigma_t^+)^{N-n}} d\cq\, 
d_3(\cq)\, v_B(t,\cq)\, \tr_{\spin_{\cq}} \,\psi_t(q,\cq) \, 
\psi_t^*(\aq,\cq) = \nonumber\\
&= I_1 +I_2 + I_3+I_4\,,\label{I1I4}
\end{align}
where $I_1+I_2+I_3$ correspond to the first term on the right hand side of \eqref{fundthm2}, and $I_4$ to the second. 

To evaluate $I_4$, note that $\partial B_t$ consists of $N-n$ facets of the form $(\Sigma_t\cap \surface)\times (\Sigma_t^+)^{N-n-1}$ and permutations thereof; on the first facet,
\be
v_B(t,\cq) = v_{\surface}(t,\cq_1)\,,
\ee
where $v_{\surface}$ is the speed at which $\surface_t=\Sigma_t\cap\surface$, regarded as a surface in coordinate space, moves in the normal direction. Exploiting the permutation symmetry of $\psi$ and
\be
(N-n)\binom{N}{n}=
(n+1)\binom{N}{n+1}
\ee
for $n<N$,
we thus obtain
\begin{align}
I_4
&= (N-n)\binom{N}{n} \int\limits_{\surface_t} dx \, d_3(x)\, 
v_{\surface}(t,x) \,\times \nonumber\\
&\quad \times \tr_{\spin_x}
\int\limits_{(\Sigma_t^+)^{N-n-1}} d\cq \,d_3(\cq)\,
\tr_{\spin_{\cq}}\, 
\psi_t(q,x,\cq) \, \psi_t^*(\aq,x,\cq)=\nonumber\\
&=
(n+1)\int\limits_{\surface_t} dx \, d_3(x)\, v_{\surface}(t,x)\, \tr_{\spin_x}\,
\dm^-(q,x;\aq,x)\,, \label{I4}
\end{align}
which is also true for $n=N$ since then both sides vanish.

The Hamiltonian in $I_2$ has two contributions, one acting on $q$ and the other on $\cq$:
\be
\hat H_t = \hat H_t(q) + \hat H_t(\cq)\,.
\ee
Correspondingly, we split
\be\label{I2}
I_2=I_2(q) + I_2(\cq)\,.
\ee
The same calculation that leads to the continuity equation for the probability density and current of the many-particle Dirac equation also shows (when applied only to $\cq$) that the integrands of $I_1$, $I_2(\cq)$, and $I_3(\cq)$ together equal
\be
-\sum_{i=1}^{3\#\cq} \partial_i J_i(q,\aq,\cq) :=
-\sum_{k=1}^{\#\cq}\sum_{\mu=1}^{3} \partial_{k,\mu}\Bigl(
d_3(\cq\setminus \cq_k) \, d_4(\cq_k) \tr_{\spin_{\cq}}\bigl( 
\psi_t(q,\cq) \, \psi^*_t(\aq,\cq)\, \alpha^\mu(\cq_k)  \bigr)\Bigr) \,,
\ee
where the $\alpha_i$ act only on the spin indices of particles belonging to $\cq$, not $q$ or $\aq$. By the Ostrogradski--Gauss integral theorem, the integral of the divergence is the flux across the surface, so that, with $\vec{n}(\cq)$ the outward-pointing unit normal vector (in coordinate space) on the surface $\partial B_t$ at $\cq$,
\begin{align}
I_1 + I_2(\cq) + I_3(\cq) 
&= -\binom{N}{n} \int_{\partial B_t} d\cq \, 
\vec{n}(\cq) \cdot \vec{J}(q,\aq,\cq) =\nonumber\\
&= -(N-n) \binom{N}{n} \int\limits_{\surface_t} dx \, d_4(x)\:\times\nonumber\\
&\quad \times \tr_{\spin_x}  
\int\limits_{(\Sigma_t^+)^{N-n-1}} d\cq \,d_3(\cq)\,\tr_{\spin_{\cq}}\,
\psi_t(q,x,\cq) \,\psi^*_t(\aq,x,\cq)\, \alpha^\perp(x) =\nonumber\\
&= -(n+1) \int\limits_{\surface_t} dx \, d_4(x)\: \tr_{\spin_x}  \,
\dm^-_t(q,x;\aq,x)\, \alpha^\perp(x) \,. \label{Icq}
\end{align}
Here we used that on the first facet of $\partial B_t$, i.e., on $\surface_t\times (\Sigma_t^+)^{N-n-1}$, the vector $\vec{n}(\cq)$ has $3(N-n)$ components of which only the first 3 can be nonzero, which form the unit normal vector on $\surface_t$ in coordinate 3-space.

In the term $I_2(q)$, the Hamiltonian $\hat H_t(q)$ acts on a variable that is not integrated over, and thus can be exchanged with the integration, which leads to
\be\label{I2q}
I_2(q) = -\tfrac{i}{\hbar}\hat H_t(q) \dm_t^-
\ee
and likewise
\be\label{I3q}
I_3(q) = \tfrac{i}{\hbar} \dm_t^-\hat H_t(\aq)\,.
\ee

Putting together \eqref{I1I4}, \eqref{I4}, \eqref{I2}, \eqref{Icq}, \eqref{I2q}, and \eqref{I3q}, we obtain that
\be\label{dm-evol}
\frac{\partial\dm_t^-}{\partial t} 
=\tfrac{i}{\hbar} [\dm_t^-,\hat H_t] + \tilde\Lop \dm_t^-
\ee
with
\be\label{dm-Lop1}
\tilde\Lop \dm_t^-(q;\aq) =  (\#q+1)\,\delta_{\#q,\#\aq}
\int\limits_{\surface_t} dx \,  \tr_{\spin_x}
\Bigl(\dm^-_t(q,x;\aq,x) \, \bigl(d_3(x)\,v_{\surface}(t,x)\,I- 
d_4(x)\,\alpha^\perp(x)\bigr) \Bigr)\,.
\ee
If we drop the assumption that $\psi$ lies in the $N$-particle sector of Fock space, we obtain instead of \eqref{dmint} that, for $q,\aq\in\conf(\Sigma^-)$, 
\be\label{dmint2}
\dm^-(q;\aq) = 
\int\limits_{\conf(\Sigma^+)} d\cq\, d_3(\cq) \:
\binom{\#q+\#\cq}{\#q}^{1/2} \binom{\#\aq+\#\cq}{\#\aq}^{1/2}
\tr_{\spin_{\cq}}\bigl(\psi(q,\cq) \, \psi^*(\aq,\cq)\bigr)
\ee
and instead of \eqref{dm-Lop1} that
\begin{multline}\label{dm-Lop}
\tilde\Lop \dm_t^-(q;\aq) = 
\sqrt{(\#q+1)(\#\aq+1)} \:\times\\
\times \:\int\limits_{\surface_t} dx \,  \tr_{\spin_x}
\Bigl(\dm^-_t(q,x;\aq,x) \, \bigl(d_3(x)\,v_{\surface}(t,x)\,I- 
d_4(x)\,\alpha^\perp(x)\bigr) \Bigr)\,.
\end{multline}
This equation agrees with \eqref{Lop6} and thus with \eqref{Lop}, except that the need for a limit $y\to x$ evaporates as the point $x$ is now not singular.

\bigskip

In our Minkowski example around \eqref{exS}, $\Sigma_t$ is coordinatized as $\RRR^3$; the Riemannian metric on $\Sigma_t$ is the flat Euclidean metric on $\RRR^3$; $d_3(x)=1$ and $d_4(x)=1$ everywhere; $\spin_x=\CCC^4$; $T(\xi)= T_0+\eta \cdot \xi$ for the appropriate $T_0\in \RRR$ and $\eta\in \RRR^3$. We choose the Lorentz frame in $\st$ such that $T_0=0$ and $\eta=(0,0,\kappa)$ with constant $0<\kappa<1$; such a choice is possible within those coordinate systems for which the time coordinate is constant on the time leaves. Now $\surface_t$, which is a 2-dimensional affine plane in $\RRR^3$, has the particularly simple form $\surface_t  = \{\xi\in\RRR^3: \xi_3 = t/\kappa\}$; the gradient of $T$ in $\RRR^3$ is $\grad T(\xi) = \eta$; the speed at which $\surface_t$ moves is $v_{\surface}(t,\xi) = 1/|\grad T(\xi)| =1/|\eta|=1/\kappa$. Thus,
\begin{align}
\tilde\Lop \dm_t(q;\aq)
&= \sqrt{(\#q+1)(\#\aq+1)}
\int\limits_{\RRR^2} d^2x \, \tr_{\spin_x}
\Bigl(\dm^-_t(q,(x,\tfrac{t}{\kappa});\aq,(x,\tfrac{t}{\kappa})) \, \bigl(\tfrac{1}{\kappa}- \alpha_{3,\spin_x}\bigr) \Bigr)\,,
\end{align}
where
\be
\alpha_{3,\spin_x} = 
\begin{pmatrix}
0&0&1&0\\
0&0&0&-1\\
1&0&0&0\\
0&-1&0&0
\end{pmatrix}
\ee
is the standard from of the third Dirac alpha matrix, acting on the spin index associated with the particle at $(x,t/\kappa)$.

\subsection{Absorbing Hypersurfaces}
\label{sec:absorb}

We now present another derivation of our Bohm-type dynamics in the presence of a singularity, along the following lines. We first consider, instead of the singularity $\sing$, a spacelike hypersurface $\surface$, and set up a version of Bohmian mechanics in which every particle gets annihilated when hitting $\surface$; we call this model \emph{Bohmian mechanics with absorption at $\surface$}. Then we let $\surface$ approach a singularity $\sing$ and argue that Bohmian mechanics with absorption at (the hypersurface) $\surface$ converges to Bohmian mechanics with absorption at (the singularity) $\sing$, i.e., to the theory described in Section~\ref{sec:overviewneweq}.

The definition of Bohmian mechanics with absorption at the hypersurface $\surface$ is natural and straightforward. Let $\psi$ be the wave function as it would evolve without absorption, let $\tilde\dens$ and $\tilde\jmath$ be the probability distribution and the probability current that $\psi$ defines on any $\Sigma^N$, let $\dens$ and $j$ be the appropriate marginals of $\tilde\dens$ and $\tilde\jmath$ on $\Sigma^-$, and use them to define Bohmian trajectories by means of the equation of motion \eqref{Bohmj}, $dQ/dt=j/p$. Observe that $\dens$ and $j$ can be obtained directly, without reference to $\psi$, from $\dm^-$, which is defined as in the previous subsection as the partial trace of $\pr{\psi}$ over the future of $\surface$ and evolves according to the quasi-Lindblad equation \eqref{dm-evol} with \eqref{dm-Lop}. In Bohmian mechanics with absorption at $\surface$, we do not mention $\psi$ but take a density matrix $\dm_t=\dm_t^-$ to be fundamental and to evolve according to the quasi-Lindblad equation \eqref{dm-evol} with \eqref{dm-Lop}, and we define $\dens$, $j$, and the law of motion in terms of $\dm_t$. We now give the defining equations.

The probability density $\dens(q)$ on configuration space $\conf(\Sigma^-)$ for $\Sigma\in\foliation$ is defined in the following way. Let $\tilde{\dens}$ be the probability density that would arise from $\psi$ without absorption at $\surface$; $\tilde{\dens}$ is defined on $\conf(\Sigma)$. Using the identification
\be
\conf(A\cup B) \cong \conf(A) \times \conf(B)
\ee
if $A\cap B=\emptyset$ (consisting of re-ordering a configuration $q$ in such a way that the particles in $A$ are listed first, $q\cong (q\cap A,q\cap B)$), and ignoring $\Sigma \setminus (\Sigma^+\cup \Sigma^-)$ because it has lower dimension, the function $\tilde\dens$ can be written as $\tilde{\dens}(q,\cq)$ with $q\in\conf(\Sigma^-)$ and $\cq\in\conf(\Sigma^+)$. We have that
\be
\tilde{\dens}(q,\cq) =  d_3(q,\cq)\, \psi^*(q,\cq) \, \psi(q,\cq)\,.
\ee
Since we assume that particles vanish when hitting the hypersurface $\surface$ while other particles are not affected, we are led to
\be
\dens(q) = \int\limits_{\conf(\Sigma^+)} d\cq \, 
\binom{\#q + \#\cq}{\#q}\,\tilde{\dens}(q,\cq)\,.
\ee
This can be re-expressed in terms of $\dm^-$, see \eqref{dmint}, as
\be
\dens(q) =  d_3(q)\,\tr_{\spin_q} \dm^-(q;q)\,,
\ee
which is the density in configuration space associated with $\dm^-$ in the natural way, parallel to \eqref{densdm1}.

Likewise, the probability current vector field $j$ on configuration space is defined as follows. Let $\tilde\jmath$ be the probability current vector field obtained from $\psi$ and defined on $\conf(\Sigma)$: for $q\in(\Sigma^-)^n$ and $\cq\in(\Sigma^+)^m$,
\be
\tilde\jmath_k^\mu(q,\cq) = d_3(q\setminus q_k) \, d_3(\cq)\, d_4(q_k)\, \psi^*(q,\cq) \,\alpha^\mu(q_k) \, \psi(q,\cq)
\ee
with $k=1,\ldots,n$. 
We define $j$ to be the marginal current on $\conf(\Sigma^-)$, or, in other words, the average current given $q$; explicitly, for $q\in (\Sigma^-)^n$,
\be
j_k^\mu(q) = \int\limits_{\conf(\Sigma^+)} d\cq \, \tilde\jmath_k^\mu(q,\cq)\,.
\ee
As a consequence,
\be
j^\mu_k(q) = d_3(q\setminus q_k)\, d_4(q_k)\,
\tr_{\spin_q} \bigl( \dm^-(q;q) \, \alpha^\mu(q_k)\bigr)\,.
\ee

Now take \eqref{dm-evol} with \eqref{dm-Lop} to be the fundamental evolution law of a density matrix $\dm_t=\dm_t^-$. For these equations to define an evolution law, we need that $\surface$ is spacelike,\footnote{In case $\surface$ is not spacelike, as long as it divides $\st$ into two connected components $\st^\pm$, i.e., $\st\setminus\surface=\st^+\cup\st^-$, then, for $\Sigma^\pm_t=\Sigma_t \cap \st^\pm$, the derivation of \eqref{dm-evol} with \eqref{dm-Lop1} remains valid, but does not uniquely determine the evolution of $\dm_t^-$ from an initial datum $\dm_{t_0}^-$.} which we always assumed, and the following locality property of the one-particle Dirac equation:\footnote{Most other relativistic wave equations share this locality property, in particular the Maxwell and Weyl equations, but not the first-order Klein--Gordon equation $i\partial_t \psi = \sqrt{m^2-\Laplace}\psi$.} Suppose $\phi$ is a one-particle wave function; in order to predict $\phi(x)$ for some space-time point $x$, one needs only the initial data in the past of $x$, i.e., $\phi|_{\Sigma_0 \cap J^-(x)}$ if $\Sigma_0$ is the surface on which the initial data are specified. In particular, if $x\in J^-(\surface)$ then $\Sigma_0\cap J^-(x) \subseteq J^-(\surface)$. As a consequence, the time evolution equation \eqref{dm-evol} can be solved on $J^-(\surface)$ without ever computing $\psi$ at any configuration containing any point from $J^+(\surface)$, and without knowing the electromagnetic field or the metric at any point of $J^+(\surface)$; indeed, the evolution equation \eqref{dm-evol} can be solved using initial data $\dm^-_{0}$ on a surface $\Sigma_0$ that already intersects $\surface$. 

Having obtained the density matrix $\dm_t=\dm_t^-$, let it define $\dens$ and $j$ as above, and let the particles move  according to \eqref{Bohmj}, whenever none of the particles hits the absorbing surface $\surface$. As soon as a particle hits $\surface$, that particle gets annihilated and removed from the configuration, as in \eqref{Qjump} with $\sing$ replaced by $\surface$.

Let us turn again, in more detail, to the difference between \emph{ignoring} and \emph{absorbing} particles behind $\surface$. While the density $\dens$ and the current $j$ on $\Sigma^-$ are merely the marginals of the density $\tilde{\dens}$ and the current $\tilde\jmath$ on $\Sigma$, the Bohm-type trajectory $Q_t$ obtained from $\dens$ and $j$ by \eqref{Bohmj} are usually very different from the trajectory $\tilde{Q}_t$ obtained from $\tilde{\dens}$ and $\tilde\jmath$ (and thus from $\psi$): Whereas the velocity of particle 1 in the configuration $\tilde{Q}_t$ may depend on the position of particle 2 that has already crossed $\surface$, the velocity of particle 1 in $Q_t$ does not so depend; after all, particle 2 does not exist any more after hitting $\surface$, and its position is therefore not defined. Instead, the velocity of particle 1 in $Q_t$ is equal, when expressed in terms of $\psi$, to the \emph{average} (of the velocity of particle 1 in $\tilde{Q}_t$) over all positions that particle 2 might assume behind $\surface$. Hence, while on one mathematical level---the level of wave functions, density matrices, probability densities and currents---it may seem like our construction merely involves \emph{ignoring} the particles behind $\surface$, on another mathematical level---the level of the trajectories---the absorption of particles at $\surface$ has an effect on the other particles that mere ignoring would not have.

The evolution of the Bohmian particles just defined is equivariant, i.e., if the configuration $Q_{t}$ on $\Sigma_{t}$ is random with distribution density $\dens^{\dm_{t}}$ then for every $s>t$, $Q_s$ has distribution density $\dens^{\dm_s}$. This can be shown with the same argument as used for \eqref{conti3}, \eqref{conti4}, and \eqref{Lop5}.


Now we assume the existence of a spacelike singularity $\sing$ in the future of $\surface$. Then the evolution of $\psi$, from which we obtained that of $\dm^-$ by a partial trace in Section~\ref{sec:reduced}, is not defined any more, but the evolution \eqref{dm-evol} of $\dm^-$ is still defined because of the locality property mentioned above. As we let the spacelike hypersurface $\surface$ approach $\sing$, \eqref{dm-evol} with \eqref{dm-Lop} formally converges to \eqref{dmevol} with \eqref{Lop}, while the laws for the Bohmian configuration remain unchanged. We thus obtain the Bohm-type evolution as in \eqref{dmevol}--\eqref{Qjump} in a different way, as a limit of the evolution with absorption at a hypersurface.

\subsection{Abstract Mathematical Structure of the Time Evolution}
\label{sec:abstract}

The time evolution we are considering, as summarized by \eqref{dmevol}, does not fit into the usual categories of quantum mechanical time evolution, as it corresponds neither to a unitary one-parameter group $U_t$ on a fixed Hilbert space $\Hilbert$ nor to a quantum dynamical semigroup (as would arise from a Lindblad equation) on $\Hilbert$. Rather, with every time leaf $\Sigma_t$ there is associated a Hilbert space $\Hilbert_t$, and the time evolution from $\Sigma_s$ to $\Sigma_t$ corresponds to a superoperator
\be
\superop_s^t:TRCL(\Hilbert_s)\to TRCL(\Hilbert_t)\,,
\ee
where $TRCL(\Hilbert)$ means the trace class of $\Hilbert$ (roughly, the set of operators $\Hilbert\to\Hilbert$ with finite trace). 

Let me elaborate a bit on the fact that $\Hilbert_s$ and $\Hilbert_t$ are not the same space: It is always the case in relativity that the time evolution takes place from one hypersurface $\Sigma_s$ to another $\Sigma_t$, and therefore that $\Hilbert_s$, containing functions of configurations in $\Sigma_s$, is different from $\Hilbert_t$, containing functions of configurations in $\Sigma_t$. On the other hand, one may often seek a way of identifying $\Sigma_s$ and $\Sigma_t$, for example by means of coordinates. In the presence of a spacelike singularity, however, coordinates $x^\mu$ for which $x^0$ is timelike and $x^1,x^2,x^3$ are spacelike may have the feature, like the Kruskal coordinates in the Schwarzschild space-time, that $\Sigma_s$ and $\Sigma_t$ correspond to different subsets of coordinate 3-space, so that the coordinates do not provide a diffeomorphism $\Sigma_s\to\Sigma_t$.  

So we are forced, more than ever, to regard $\Hilbert_s$ and $\Hilbert_t$ as different spaces. As a consequence, the time evolution superoperator $\superop_s^t$ must have two indices, indicating the initial time $s$ and the final time $t$. And as a consequence of \emph{that}, they cannot form a semigroup, since a semigroup $\{g_t\}$ requires that $g_t g_u = g_{t+u}$. Instead, the appropriate notion of time evolution for our purpose is that of \textit{cocycle}, a notion taken from the theory of random dynamical systems \cite{random} and designed for describing the evolution in the presence of noise or time-dependent external fields. 


We define a \emph{quantum dynamical cocycle} to be a 2-parameter family of mappings
\be
\superop_s^t:TRCL(\Hilbert_s)\to TRCL(\Hilbert_t)
\ee
for $0\leq s\leq t$ such that
\begin{itemize}
\item $\superop_s^t$ is $\CCC$-linear, completely positive, and trace-preserving; 
\item $\superop_t^u \superop_s^t = \superop_s^u$ for $0\leq s \leq t \leq u$;
\item $\superop_t^t(\dm) = \dm$ for every $\dm \in TRCL(\Hilbert_t)$;
\end{itemize}

It is plausible that equations \eqref{dmevol} and \eqref{Lop} define a quantum dynamical cocycle $\superop_s^t$: To begin with, \eqref{dmevol} is linear in $\dm_t$. To argue that $\superop_s^t$ is completely positive, we note that the composition of completely positive superoperators is completely positive; regarding $\superop_s^t$ as the composition of many $\superop_u^{u+du} = I + \tfrac{i}{\hbar} [\cdot,\hat H_t]dt + \Lop\,dt$, we need that both the Hamiltonian evolution and $\Lop$ are completely positive, which is plausible. The conservation of $\tr \dm_t$ corresponds to the conservation of total probability, which is exactly what $\Lop$ was designed for.

Another requirement that one may wish to add to the definition is that $\superop_s^t$ depend continuously on $s$ and $t$, but for this one needs a topology on the bundle $\cup_t \Hilbert_t$, which we have not defined yet. More importantly, when requiring continuity in $s$ and $t$ then the definition fails to cover the cases in which $\Sigma_t\cap \sing$ is 3-dimensional, see \eqref{chunk}.

Note that a quantum dynamical semigroup forms a special case of a quantum dynamical cocycle, in which all Hilbert spaces $\Hilbert_t$ are identified with one fixed space $\Hilbert$ and $\superop_s^t$ depends only on the time difference $t-s$. 

Another special case arises for a space-time without spacelike singularities, in which the quantum-mechanical time evolution is unitary. Then $\superop_s^t(\dm) = \hat U_s^t \dm \hat U_t^s$, where the $\hat U_s^t:\Hilbert_s \to \Hilbert_t$ are unitary isomorphisms implementing the unitary evolution of the wave function according to $\psi_t = \hat U_s^t \psi_s$; they satisfy $\hat U_t^t \psi = \psi$ and $\hat U_t^u \hat U_s^t = \hat U_s^u$ for all $s,t,u\in\RRR$. In this case we call $(\superop_s^t)_{0\leq s \leq t}$ a \textit{unitary cocycle}. As an example of a unitary time evolution that does not correspond to a unitary 1-parameter group, but can be represented as a unitary cocycle, consider a time-dependent Hamiltonian $\hat H_t$ on a fixed Hilbert space, as in non-relativistic quantum mechanics with time-dependent external fields; then (for bounded $\hat H_t$)
\be
\hat U_s^t = \mathcal{T} \exp\Bigl(-i\hbar^{-1}\int_s^t \hat H_\tau \, d\tau\Bigr)
\ee
with $\mathcal{T}\exp$ the time-ordered exponential; in other words, $\hat U_s^t$ is the solution of the initial-value problem
\be
\frac{\partial \hat U_s^t}{\partial t} = -\tfrac{i}{\hbar} \hat H_t \hat U_s^t \,, \quad
\hat U_s^s = I\,.
\ee
In order to define a Hamiltonian operator, or infinitesimal generator, for more general unitary cocycles for which the $\Hilbert_t$ are not identified, we need a differentiable structure on the bundle $\cup_t\Hilbert_t$ of Hilbert spaces. We begin with outlining an infinite-dimensional analog of the concept of a vector bundle.

By a \emph{Hilbert bundle} over the interval $(t_1,t_2)\subseteq \RRR$ we mean a family of Hilbert spaces $\Hilbert_t$, $t\in(t_1,t_2)$, together with an equivalence class of bundle maps; a \emph{bundle map} or \emph{trivialization} is a family $\hat V_t:\Hilbert_t\to\Hilbert$ of unitary isomorphisms to some fixed Hilbert space $\Hilbert$; two bundle maps $\hat V_t$, $\hat V'_t:\Hilbert_t\to\Hilbert'$ are \emph{equivalent} if and only if for every $\psi\in \Hilbert$ and one (and thus every) unitary isomorphism $\hat W:\Hilbert' \to\Hilbert$, the curve $t\mapsto \hat W \hat V'_t \hat V^{-1}_t\psi$ in $\Hilbert$ is $C^\infty$. We simply write $\cup_t \Hilbert_t$ to denote the bundle (with the understanding that $\Hilbert_t \cap \Hilbert_s = \emptyset$ for $t\neq s$), and $\hat V: \cup_t\Hilbert_t \to (t_1,t_2)\times \Hilbert$ to denote the bundle map consisting of the $\hat V_t:\Hilbert_t \to \Hilbert$.

The definition of a Hilbert bundle over manifolds other than intervals would involve \emph{local bundle maps} (or \emph{local trivializations}) and requirements on the \emph{transition functions} between two local bundle maps concerning smoothness and consistency. The 1-dimensional case, all we need here, is particularly simple. In our setting, the base manifold of the Hilbert bundle is really the time foliation $\foliation$, regarded as a 1-dimensional manifold containing one point representing each $\Sigma_t$; since we parameterize the time leaves $\Sigma_t$ by real numbers $t$ anyhow, we can afford to regard the base manifold as simply an interval $(t_1,t_2)$ of $t$-values.

Now we want to define an extension to Hilbert bundles of the concept of a \emph{connection} on a vector bundle. A connection essentially amounts to an identification between the fiber spaces over infinitesimally neighboring base points, leading to a path-dependent identification between the fiber space over any two base points, known as \emph{parallel transport}. Relative to a choice of bases in the fiber spaces, the connection can be expressed in terms of connection coefficients $\Gamma^a_{\:\:b\sigma}$, the best-known example of which are the Christoffel symbols, the coefficients of the connection naturally associated with a Lorentzian (or Riemannian) metric. Of the three indices $a,b,\sigma$, the last one refers to the base manifold (for the Christoffel symbols, space-time; for us, the time axis as represented by $\foliation$), while $a$ and $b$ refer to the fiber space (for Christoffel symbols, again the tangent space to space-time; for us, $\Hilbert_t$). Since for us, the base manifold is one-dimensional, we can drop the index $\sigma$, taking $\partial/\partial t$ as the basis vector along the base manifold; what remains is $\Gamma^a_{\:\:b}$, an operator on $\Hilbert_t$, which we write as $-i\hat H_t/\hbar$; if parallel transport respects inner products, then $\Gamma^a_{\:\:b}$ must be skew-adjoint, or $\hat H_t$ self-adjoint.

We can thus define a \emph{connection on a Hilbert bundle $\cup_t\Hilbert_t$ over $(t_1,t_2)$} to be given by an equivalence class of triples $(\Hilbert,\hat V, (\hat H_t)_{t\in(t_1,t_2)})$, where $\Hilbert$ is a Hilbert space, $\hat V:\cup_t \Hilbert_t\to(t_1,t_2)\times\Hilbert$ is a bundle map, and $(\hat H_t)_{t\in(t_1,t_2)}$ is a 1-parameter family of operators $\hat H_t$ on $\Hilbert$. If every $\hat H_t$ is self-adjoint then we call the connection \emph{unitary}. Two triples $(\Hilbert,\hat V, (\hat H_t))$ and $(\Hilbert', \hat V', (\hat H'_t))$ are considered equivalent if and only if 
\be\label{Htransform}
\hat V_t (\hat V'_t)^{-1} \hat H'_t \hat V'_t \hat V^{-1}_t = 
\hat H_t 
-i\hbar \frac{d}{dt} \Bigl(\hat V_t (\hat V'_t)^{-1} \Bigr) \hat V'_t  \hat V_t^{-1} 
\ee
where the derivative is understood as a strong limit. To understand this definition, think of the connection as defining a parallel transport operator $\hat U_s^t:\Hilbert_s\to\Hilbert_t$ by means of
\be
\hat U_s^t = \hat V_t^{-1} \mathcal{T} \exp\Bigl( -i\hbar^{-1}\int_s^t \hat H_u \, du \Bigr) \hat V_s\,.
\ee
Since $\hat U_s^t$ should be independent of the bundle map $\hat V$, $\hat H_t$ has to transform in the appropriate way, and a short calculation shows that \eqref{Htransform} is the transformation law. 

As an example of a Hilbert bundle and a connection, consider a single Dirac particle on a singularity-free space-time $\st$ with global coordinates $(x^0,\ldots,x^3) : \st \to \RRR^4$, forming a diffeomorphism, such that $x^0$ is timelike and $x^1,x^2,x^3$ are spacelike. 
Then $\Sigma_t=\{x^0=t\}$ is a spacelike hypersurface, and $\varphi_t = (x^1,x^2,x^3): \Sigma_t \to\RRR^3$ is a diffeomorphism. The Hilbert space $\Hilbert_t$ consists of measurable cross-sections of the spin bundle $\cup_{x\in \Sigma_t} \spin_x$ that are square-integrable relative to the Riemannian 3-volume measure on $\Sigma_t$ and relative to the inner product on $\spin_x$ associated with the unit normal vector on $\Sigma_t$ at $x$. Let $\Hilbert = L^2(\RRR^3,\CCC^4)$ with respect to Lebesgue measure on $\RRR^3$ and the standard inner product on $\CCC^4$. Note that the Riemannian 3-volume measure has non-constant density $d_3(x)=\sqrt{-\det g^{(3)}(x)}$ relative to the Lebesgue measure on coordinate space $\RRR^3$, so we need to compensate for that. At every $x\in\Sigma_t$ choose an orthonormal basis $\tilde b_x$ of $\spin_x$ that depends smoothly on $x$. Obtain $b_x$ from $\tilde b_x$ by scaling each basis vector by $\sqrt{d_3(x)}$. Then define $\hat V_t$ by
\be
 (\hat V_t \psi)_s(q) = b_{x,s}^* \psi(x)\Big|_{x=\varphi_t^{-1}(q)}\,,
\ee
where $q\in\RRR^3$ is a point in coordinate space, $s\in\{1,2,3,4\}$ a spin index, $b_{x,s}$ the $s$-th element of $b_x$, and $\phi^* \psi$ the inner product in $\spin_x$. To see that $\hat V_t$ is unitary, note that
\[
\scp{\hat V_t \phi}{\hat V_t \psi} = 
\int_{\RRR^3} dq \, \sum_{s} \phi^*(x)  b_{x,s}\, b_{x,s}^* \psi(x) \Big|_{x=\varphi_t^{-1}(q)}
\]
\[
= \int_{\RRR^3} dq \, d_3(x)\,  \phi^*(x) \psi(x) \Big|_{x=\varphi_t^{-1}(q)}
=\scp{\phi}{\psi}_{\Hilbert_t}\,,
\]
and that $\hat V_t$ is clearly surjective. The Dirac equation defines a unitary time evolution $\hat U_s^t:\Sigma_s \to \Sigma_t$, corresponding to a unitary connection on $\cup_t\Hilbert_t$, expressed in the coordinates $x^\mu$ as a time-dependent self-adjoint Hamiltonian $\hat H_t$.

In the presence of a future singularity, the $\Hilbert_t$ still form a Hilbert bundle. For example, in the Schwarzschild space-time with the $t'$-foliation for $t'> 0$ and $\Hilbert_{t'} = \Gamma_\eith\bigl(L^2(\Sigma_{t'},\spin|_{\Sigma_{t'}})\bigr)$, where $\Gamma_\eith$ means either the bosonic or the fermionic Fock space, we can set $\Hilbert = \Gamma_\eith(L^2(\RRR\times\SSS^2,\CCC^4))$. For $0<t'<2M$, we use the coordinates as before to identify $\Sigma_{t'}$ with $\RRR\times\SSS^2$. For $t'\geq 2M$, $\Sigma_{t'}$ has two connected components, $\Sigma_{t'}^{(+)}$ with $x'>\sqrt{t^{\prime 2} -2M}$ and $\Sigma_{t'}^{(-)}$ with $x'<-\sqrt{t^{\prime 2} -2M}$; by replacing the $x'$ coordinate on $\Sigma_{t'}^{(\pm)}$ with $x''=x'\mp \sqrt{t^{\prime 2}-2M}$, we identify $\Sigma_{t'}$ with $(\RRR\setminus\{0\})\times\SSS^2$; the one point missing from $\RRR$ does not affect the $L^2$ space. Then the bundle map $\hat V$ can be defined as before.

In the presence of a future singularity, the Dirac equation fails to define a unitary evolution, but it still defines a time evolution for a wave function $\phi_t\in\Hilbert_{\Sigma_t}$, one for which $\|\phi_t\|$ decreases with increasing $t$. For example, in the setting of Section~\ref{sec:absorb}, involving instead of a singularity an absorbing spacelike hypersurface $\surface$, $\phi_t$ is just the restriction of $\psi_t$ (the wave function in the absence of the absorbing hypersurface) to $\conf(\Sigma_t^-)$, i.e., $\phi_t$ is $\psi_t$ evaluated only at configurations for which all particles are located in $\Sigma_t^-$, i.e., in the past of the hypersurface $\surface$; correspondingly, $\|\phi_t\|^2$ is the probability that no particle has hit $\surface$ up to time $t$. In other words, in the presence of a future singularity the Dirac equation still defines a connection on $\cup_t \Hilbert_t$, the \emph{Dirac connection}, but it is not unitary; correspondingly, the Dirac Hamiltonian is not self-adjoint.

In terms of the Dirac connection, the evolution \eqref{dmevol} can be expressed as
\be
\nabla \dm_t = \Lop \dm_t\,,
\ee
where $\nabla$ is the covariant derivative operator associated with the Dirac connection (and applied along the vector field $\partial/\partial t$ on the time axis), and $\dm_t$ is now a cross-section of the Banach space bundle $\cup_t TRCL(\Hilbert_t)$ arising from the Hilbert bundle $\cup_t \Hilbert_t$.

\subsection{Remarks}

\textit{A Chunk of Singularity. }
In equation \eqref{dmevol} we assumed that $\Sigma_t\cap \sing$ is 2-dimensional; if it is 3-dimensional for a particular $t_0$ then $\dm_t$ should have a discontinuity as a function of $t$ at $t_0$, according to
\be\label{chunk}
\dm_{t_0+}(q;\aq) = \dm_{t_0-}(q;\aq) + 
\int\limits_{\conf(\Sigma_{t_0}\cap \sing)} d\cq \, 
\tr_{\spin_{\cq}}\, \dm_{t_0-}(q,\cq;\aq,\cq)
\ee
for $q,\aq\in \conf(\Sigma_{t_0}\setminus \sing)$.

\bigskip

\textit{Non-Smooth Singularities. }
Penrose \cite{Pen79} has suggested that the future singularities arising from gravitational collapse may be rather irregular, and this further suggests that they may in fact be non-smooth. That is, after a conformal transformation the singularity may correspond to an achronal surface that is not smooth but merely continuous. In our discussion so far we assumed smoothness, but it seems plausible that this assumption is not needed.

The reason for believing this is that smoothness is not needed when considering an absorbing spacelike hypersurface $\surface$ instead of a singularity $\sing$. Suppose $\surface$ is merely continuous. 
If $\surface\cap\Sigma_t$ is a null set in every $\Sigma_t$, then $\Sigma_t^\pm$ and $\dm_t^\pm$ are still well defined, since their definition did not involve differentiation. That $\dm_t^-$ is well defined is all we needed to show.

To be sure, $\dm_t^-(q,\aq)$ may fail to be differentiable with respect to $t$ as a consequence of the lack of smoothness in the $t$-dependence of $\Sigma_t^+$; however, $\dm_t^-$ may also fail to be differentiable with respect to $t$ if any of its eigenfunctions is not contained in the domain of $\hat H_t$. After all, the Schr\"odinger equation
\be
i\hbar \frac{d\psi_t}{dt} = \hat H_t \psi_t
\ee
holds literally only for $\psi_t$ in the domain of $\hat H_t$; if $\psi_{t_0}$ lies outside the domain of $\hat H_{t_0}$ then $t\mapsto \psi_t$ is not differentiable at $t_0$. To sum up, even though \eqref{dm-evol} cannot be expected to hold literally, the evolution of $\dm_t^-$ should exist for non-smooth $\surface$.

\bigskip

\textit{Objections. }
\label{sec:objection}
Banks, Susskind, and Peskin \cite{BSP84} have argued, in response to Hawking's \cite{Haw82} proposal that the fundamental time evolution might transform pure to mixed states, that such an evolution would have to either allow superluminal signalling or violate the conservation of energy-momentum. Indeed, in our model energy-momentum is not conserved, as the energy-momentum of a particle hitting the singularity gets lost; it should be accounted for by a suitable change in the space-time geometry, but our model does not do that. On the other hand, our model does not allow superluminal signalling. This is most directly conveyed by the parallel between the evolution equations of the density matrix $\dm_t$ in the presence of a future singularity and the density matrix $\dm_t^-=\tr_+ \pr{\psi}$ as in \eqref{dm-def} obtained by tracing out the future of a spacelike hypersurface in a non-singular space-time (see Section~\ref{sec:absorb}), and interpreting the operation of tracing out as merely ignoring part of the information encoded in $\psi$. It is known that the unitary evolution of $\psi$ does not allow superluminal signalling, and rather clear that ignoring some information 
cannot create a possibility of superluminal signalling. Still, it would be desirable to have a carefully formulated no-signalling proof.

Maudlin \cite{Mau04} has argued that the pure-to-mixed evolution be an artifact of considering the wrong spacelike hypersurfaces. If $\Sigma_s$, he argued, is a Cauchy surface 
and $\Sigma_t$ is not, for example if they are hypersurfaces of constant $t'$ in the Schwarzschild space-time with $-\sqrt{2M}<s<\sqrt{2M}$ and $t>\sqrt{2M}$, then it is no wonder that the evolution from $\Sigma_s$ to $\Sigma_t$ is pure-to-mixed: after all, if $\Sigma_t$ is not a Cauchy surface then it is not adequate for describing initial data of the evolution. Thus, the pure-to-mixed evolution does not mean that anything is unusual about the evolution but that some hypersurfaces are inadequate. For example, the hypersurface $\{t=0,x>0\}$ in Minkowski space-time is spacelike but not Cauchy---it is too small---and so the quantum state associated with it is the density matrix arising from the wave function on $\{t=0\}$ by tracing out the degrees of freedom associated with $x\leq 0$. If one sticks to Cauchy surfaces, then the evolution remains unitary, and black hole evaporation never occurs. However, this argument does not work in Bohmian mechanics as it ignores the role of the time foliation. As mentioned at the end of Section~\ref{sec:motivation}, the time leaves $\Sigma\in\foliation$ may not be Cauchy surfaces but instead border on the singularity.

\bigskip 

\textit{Black Hole Evaporation. }
There is no consensus in the literature about whether information is lost during black hole evaporation, i.e., whether unitarity is violated. 
While in our model unitarity is indeed violated, this does not support conclusions about black hole evaporation, for two reasons: First, we assumed that the gravitational field can be described by a classical Lorentzian geometry with a spacelike singularity. This assumption we might be violated, as it may be necessary to apply a quantum gravity theory, and it can be questioned whether then any singularity will actually arise. Second, our model ignores any back reaction of the particles on the space-time geometry. It is clear that a black hole will grow in mass when swallowing particles, so the quantum state of the gravitational field should be affected by the infalling particle and might store the information lost from the quantum state of the matter, leading to unitarity of the full evolution of both matter and the gravitational field.

\section{Past Spacelike Singularities}
\label{sec:past}

Our method of studying past singularities is to postulate reversibility of the fundamental laws of the theory. In this way, the laws we already have for future singularities determine the laws for past singularities.

The obvious fact about past singularities is that no future-directed timelike curve can end there. Thus, a past singularity can emit but not absorb particles. Examples of past singularities include the set $\sing_2$ in Schwarzschild space-time (see Section~\ref{sec:schwarz}) corresponding to a white hole and the big bang singularity in Friedmann--Robertson--Walker space-times \cite{HE73}.

\subsection{Evolution of the Density Matrix}

The equation \eqref{dmevol} for evolving a density matrix to the future in the presence of a future singularity can be used, when time-reversed, for evolving a density matrix to the past in the presence of a past singularity. The time-reversed form reads
\be\label{dmpast}
\frac{\partial\dm_t}{\partial (-t)} 
=-\tfrac{i}{\hbar} [\dm_t,\hat H_t] + \Lop \dm_t
\ee
with $\Lop$ defined as in \eqref{Lop}. Compared to the equation \eqref{dmevol} for a future singularity, the term $\Lop \dm_t$ has the opposite sign. If we choose an initial wave function on a Cauchy surface (such as $\{t'=0\}$ in Schwarzschild space-time), the density matrix $\dm_t$ is defined on every time leaf $\Sigma_t$.

However, the time evolution towards the future is not well defined in the presence of a past singularity (like the time evolution towards the past in the presence of a future singularity): We have to invert a pure-to-mixed evolution, and since this evolution is many-to-one, its inverse is not unique. The problem is analogous to that of recovering a vector $\psi\in\Hilbert_1\otimes\Hilbert_2$ from its reduced density matrix $\dm_1 = \tr_2 \pr{\psi}$. Thus, the evolution of $\dm_t$ towards the future is not uniquely determined. It is not governed by a stochastic law, either, but such a law could be added as follows. A theory could provide a probability distribution $\mu_\Sigma(d\psi_\Sigma)$ for the wave function $\psi_\Sigma$ on a Cauchy hypersurface $\Sigma$, preferably in a way that does not single out any particular $\Sigma$. (For examples of probability distributions over wave functions, see \cite{GAP}.) Then any initial datum $\dm_{t_0}$ on a time leaf $\Sigma_{t_0}$ bordering on the past singularity (and thus in the past of $\Sigma$) defines a conditional distribution $\mu_\Sigma(d\psi_\Sigma|\dm_{t_0})$, concentrated on the set of those $\psi_\Sigma$'s which, when evolved backward to $\Sigma_{t_0}$, lead to $\dm_{t_0}$, and thus also defines a stochastic process $(\dm_t)_{t\geq t_0}$.

\subsection{Evolution of the Configuration}

Let us ignore the problem of finding the density matrices $\dm_t$ in the presence of a past singularity---let us suppose we are given all density matrices $\dm_t$---and focus on how to define the evolution of the configuration $Q_t$. Sticking to reversibility, we obtain, from the evolution we know in the presence of future singularities, that the particles move according to \eqref{Bohmdm}, the Bohm-type law of motion using a fundamental density matrix, with new particles created at the singularity in a stochastic way, given by \eqref{pastrate}. 

Indeed, think of the evolution in the presence of future singularities as a stochastic process, i.e., as a measure $\PPP$ on path space. This process is in fact deterministic, that is, the initial configuration $Q_{t_0}$ is random with distribution $\dens^{\dm_{t_0}}$, and the path is a function of $\dm_{t_0}$ and $Q_{t_0}$. The time reversal mapping $T$ maps every path to its time reverse, and $\PPP$ to $T_*\PPP$.\footnote{In the general relativistic context, time reversal is essentially a trivial operation, since the time reverse of a space-time, decorated with world lines, is isometric to the original, and thus physically equivalent. Still, a theory may fail to be reversible if it assumes, as we did, a time orientation.} Our claim is that $T_*\PPP$ corresponds to a Markovian stochastic process with particle creation at rate \eqref{pastrate}. Being defined by a probability distribution on path space, it obviously is a stochastic process. The Markov property follows from the determinism in the opposite time direction: If the past path is a function of the present configuration (for fixed $\dm_{t_0}$), then conditional probabilities of future events given the past path equal those given the present configuration. Since between two jumps the trajectory in configuration space is deterministic in both time directions, the only randomness concerns when to jump and where to jump. The only possible jumps are, up to permutation of the configuration,
\be
Q(t-)\to Q(t+)=(Q(t-), X)
\ee 
with $X\in \sing$, corresponding to the creation of a new particle at the past singularity. (The creation of two or more particles at the same time has probability density zero.) To determine the rate of such a jump, note that in the other time direction,
\begin{multline}
\PPP\Bigl(Q(t)\in dq\times \Sigma_t, Q(t+dt) \in dq, \text{end point} \in d^2x\Bigr) =\\
\dens(t,q,x) dq\, d^2x \, \bigl(v_\sing (t,x)-v_{x,\perp}(t,q, x)\bigr)dt\,, 
\end{multline}
and therefore
\begin{multline}
T_*\PPP\Bigl(Q(t)\in dq, Q(t+dt)\in dq \times \Sigma_t, \text{creation point} \in d^2x \Bigr) =\\
\dens(t,q,x) dq\, d^2x \, \bigl(v_\sing (t,x)-v_{x,\perp}(t,q, x)\bigr)dt\,, 
\end{multline}
which implies
\begin{align}
\sigma_t(d^2x|q)\, dt 
&=T_*\PPP\Bigl(\text{creation within time } dt 
\text{ and location } d^2x \Big| Q(t)\in dq\Bigr) \nonumber\\
& =(\#q+1)\frac{\dens(t,q,x)}{\dens(t,q)}\, d^2x \, 
\bigl(v_\sing (t,x)-v_{x,\perp}(t,q, x)\bigr)dt \nonumber\\
&= (\#q+1)\frac{\lim\limits_{y\to x, y\notin\sing}\tr_{\spin_{q,y}} \dm_t(q,y;q,y) 
\bigl(d_3(y)v_\sing (t,x)\,I -d_4(y)\alpha^\perp(y)\bigr)}
{\tr_{\spin_q} \dm_t(q;q)} d^2x\, dt\,,
\end{align}
which agrees with \eqref{pastrate}.

This evolution of the configuration, based on the combination of \eqref{Bohmdm} and \eqref{pastrate}, is equivariant, i.e., if $Q(t)$ is random with distribution density $\dens^{\dm_{t}}$ defined in \eqref{densdm1} then also for every $s>t$, $Q(s)$ has distribution density $\dens^{\dm_s}$. This follows from the fact that the time-reversed process is equivariant.

\subsection{Comparison with Bell-Type Quantum Field Theory}

Stochastic jumps of the configuration also occur in an extension of Bohmian mechanics to quantum field theory (QFT), known as \emph{Bell-type QFT} \cite{crlet, crea2B, Bell86}. Let us compare the jumps in the two theories. 

In Bell-type QFT, the jumps usually correspond to creation or annihilation of particles, while a past singularity can create but not annihilate particles (assuming that the world lines are causal). Also in Bell-type QFT, the configuration process is Markovian, with the jump rate specified by a law \cite{crea2B} that reads (after replacing the wave function in Fock space by a fundamental density matrix on Fock space)
\begin{equation} \label{Bellrate2}
     \sigma_t(d\bq|q)= \frac{[(2/\hbar) \Im \, \tr(\dm_t\, P(d\bq) H_I
     P(dq))]^+}{\tr(\dm_t\, P(dq))}
\end{equation}
with $H_I$ the interaction Hamiltonian, $x^+=\max\{x,0\}$ the positive part of $x\in\RRR$, and $P(dq)$ a positive-operator-valued measure (POVM) on configuration space serving as the configuration observable, usually
\be
P(dq) = \pr{q}\otimes I_{\spin_q} \, d_3(q) \,dq\,.
\ee
Both \eqref{pastrate} and \eqref{Bellrate2} are of the form
\be\label{rate3}
\sigma_t(d\bq|q) = 
\frac{[\Re\,\tr(\dm_t \, R(d\bq\times dq))]^+}{\tr(\dm_t \, P(dq))}
\ee
but with different operator-valued measures $R(\cdot)$ on $\conf\times\conf$: In Bell-type QFT,
\be
R(d\bq \times dq) = -\tfrac{2i}{\hbar} P(d\bq) H_I P(dq)\,,
\ee
whereas in our jump rate \eqref{pastrate},
\be\label{pastRate}
R(d\bq \times dq) 
=(\#q+1)\int\limits_{x\in\sing_t} P(dq\times d^2x)\,
w(x)\, \lim\limits_{y\to x,y\notin \sing} c_\mu(x)\,d_4(y)\,\alpha^\mu(y)
\, \delta_{(q,x)}(d\bq)
\ee
with $\delta_q$ the Dirac delta measure centered at $q$.
Of course, since \eqref{pastRate} is positive-operator-valued, in this case the operations in \eqref{rate3} of taking the real part and the positive part are trivial.

The common structure \eqref{rate3} is owed to the fact that both rate formulas can be obtained starting from the appropriate formula for the \emph{probability current}
\be
J_t(d\bq,dq) = \Re\, \tr\bigl(\dm_t \, R(d\bq\times dq)\bigr)
\ee
between volume elements $d\bq$ and $dq$ in configuration space. The minimal jump rate $\sigma_t(d\bq|q)$ compatible with this current is \eqref{rate3}. The form of the operators $R(d\bq\times dq)$ is (not uniquely determined but) suggested by the probability balance equation \eqref{conti4}, respectively the probability balance equation of QFT, which we need to agree with the probability balance equation of a jump process,
\be
\frac{\partial \dens_t}{\partial t}(q) = 
-\sum_{i=1}^{3\#q} \partial_i j_i
- \int\limits_{\bq\in \conf} \frac{J_t(d\bq,dq)}{dq} \,,
\ee
where $j$ is the current due to continuous motion and $J$ the current due to jumps.

\bigskip

\noindent\textit{Acknowledgments.} 
This research was supported by grant RFP1-06-27 from The Foundational Questions Institute (fqxi.org). I gratefully acknowledge helpful discussions with Demetrios Christodoulou (ETH Z\"urich), Detlef D\"urr (LMU M\"unchen), Felix Finster (Regensburg), Michael Kiessling (Rutgers), Tim Maudlin (Rutgers), Shadi Tahvildar-Zadeh (Rutgers), and Bassano Vacchini (Milano).


\begin{thebibliography}{29}

\bibitem{BM} Baez, J.C., and Muniain, J.P.:
	\textit{Gauge Fields, Knots, and Gravity.}
	Singapore: World Scientific (1994)

\bibitem{BSP84} Banks, T., Susskind, L., and Peskin, M.E.:
   Difficulties for the Evolution of Pure States in to Mixed States. 
   \textit{Nucl. Phys. B} \textbf{244}: 125 (1984)

\bibitem{Belldm} Bell, J.S.: De Broglie--Bohm, delayed-choice
   double-slit experiment, and density matrix. \textit{Int.\ J.\
     Quant.\ Chem.}\ \textbf{14}: 155--159 (1980). Reprinted in
   \cite{Bell87b}, p.~111.

\bibitem{Bell86} Bell, J.S.: Beables for Quantum Field Theory.
  \textit{Phys. Rep.} \textbf{137}: 49--54 (1986). Reprinted as
  chapter 19 of \cite{Bell87b}.

\bibitem{Bell87b} Bell, J.S.: \textit{Speakable and Unspeakable in
    Quantum Mechanics}. Cambridge University Press (1987)

\bibitem{Bohm52} Bohm, D.: A Suggested Interpretation of the Quantum
  Theory in Terms of ``Hidden'' Variables, I and II. \textit{Physical
  Review} \textbf{85}: 166--193 (1952)

\bibitem{cpm} 
  Choi's theorem on completely positive maps.
  In \textit{Wikipedia, the free encyclopedia} (accessed May 13, 2009)
  \url{http://en.wikipedia.org/wiki/Completely_positive_map}

\bibitem{HBD} D\"urr, D., Goldstein, S., M\"unch-Berndl, K., and Zangh\`\i, N.: 
  Hypersurface Bohm--Dirac Models. 
  \textit{Phys. Rev. A} \textbf{60}: 2729--2736 (1999). arXiv:quant-ph/9801070

\bibitem{crlet} D{\"u}rr, D., Goldstein, S., Tumulka, R.,
  and Zangh{\`{\i}}, N.: Bohmian Mechanics and Quantum Field Theory.
  \textit{Phys. Rev. Lett.} \textbf{93}: 090402 (2004).
  arXiv:quant-ph/0303156

\bibitem{dm} D{\"u}rr, D., Goldstein, S., Tumulka, R., and
  Zangh{\`{\i}}, N.: On the Role of Density Matrices in Bohmian Mechanics.
  \textit{Found. Phys.} \textbf{35}: 449--467 (2005). arXiv:quant-ph/0311127

\bibitem{crea2B} D{\"u}rr, D., Goldstein, S., Tumulka, R.,
  and Zangh{\`{\i}}, N.: Bell-Type Quantum Field Theories.
  \textit{J. Phys. A: Math. Gen.} \textbf{38}: R1--R43 (2005).
  arXiv:quant-ph/0407116

\bibitem{DGZ92} D\"urr, D., Goldstein, S., and Zangh\`\i, N.: 
  Quantum Equilibrium and the Origin of Absolute Uncertainty. 
  \textit{J. Statist. Phys.} \textbf{67}: 843--907 (1992). arXiv:quant-ph/0308039

\bibitem{GKP72} Geroch, R.P., Kronheimer, E.H., and Penrose, R.:
	Ideal Points in Space-Time.
	\textit{Proc. Roy. Soc. Lond. A} \textbf{327}: 545--567 (1972)

\bibitem{Gol01} Goldstein, S.: 
	Bohmian Mechanics. 
	In E. N. Zalta (ed.), \textit{Stanford Encyclopedia of Philosophy}, 
	published online by Stanford University (2001). 
	\url{http://plato.stanford.edu/entries/qm-bohm/}

\bibitem{GAP} Goldstein, S., Lebowitz, J.L., Tumulka, R., and Zangh\`\i, N.:
       On the Distribution of the Wave Function for System in Thermal Equilibrium.
     \textit{J. Statist. Phys.} \textbf{125}: 1193--1221 (2006).
        arXiv:quant-ph/0309021

\bibitem{GKS} Gorini, V., Kossakowski, A., and Sudarshan, E.C.G.:
	Completely positive dynamical semigroups of $N$-level systems.
	\textit{J. Math. Phys.} \textbf{17}: 821 (1976)

\bibitem{Haw76} Hawking, S.W.:
	Breakdown of predictability in gravitational collapse.
	\textit{Phys. Rev. D} \textbf{14}: 2460--2473 (1976)

\bibitem{Haw82} Hawking, S.W.:
	The unpredictability of quantum gravity.
	\textit{Commun. Math. Phys.} \textbf{87}: 395--415 (1982)

\bibitem{HE73} Hawking, S.W., and Ellis, G.F.R.:
  \textit{The large scale structure of space-time}. 
  Cambridge University Press (1973)

\bibitem{Kru60} Kruskal, M.D.:
  Maximal extension of Schwarzschild metric.
  \textit{Phys. Rev.} \textbf{119}: 1743--1745 (1960)

\bibitem{Lan72} Lang, S.: \textit{Differentiable Manifolds.}
	Reading, Mass.: Addison Wesley (1972)

\bibitem{Lin76} Lindblad, G.:
  On the generators of quantum dynamical semigroups.  
  \textit{Commun. Math. Phys.} \textbf{48}: 119--130 (1976)

\bibitem{Mau04} Maudlin, T.:
	Cauchy surfaces and black hole evaporation.
	Talk given at the conference 
	\textit{Quantum Theory Without Observers II}, Bielefeld (Germany),
	2-6 February 2004

\bibitem{MTW} Misner, C.W., Thorne, K.S., and Wheeler, J.A.:
	\textit{Gravitation}. New York: Freeman (1973)


\bibitem{Pen79} Penrose, R.: 
  Singularities and time-asymmetry.
  In S.W. Hawking and W. Israel (ed.s), \textit{General relativity: An Einstein
  centenary survey}, 581--638. Cambridge University Press (1979)

\bibitem{PR84} Penrose, R., and Rindler, W.:
  \textit{Spinors and space-time. Volume 1: Two-spinor calculus and
  relativistic fields.} Cambridge University Press (1984)

\bibitem{random} Random dynamical system. 
  In \textit{Wikipedia, the free encyclopedia} (accessed July 27, 2008)
  \url{http://en.wikipedia.org/wiki/Random_dynamical_system}

\bibitem{Sch} Schwarzschild, K.: 
   \"Uber das Gravitationsfeld eines Massenpunktes nach der
   Einstein'schen Theorie.
   \textit{Sitzungsberichte der K\"oniglich Preu\ss ischen Akademie
   der Wissenschaften} \textbf{1}: 189--196 (1916)

\bibitem{3forms} Tumulka, R.:
  \textit{Closed 3-Forms and Random Worldlines}. 
  Ph. D. thesis, Mathematics Institute, Ludwig-Maximilians-Universit\"at, 
  M\"unchen, Germany (2001). 
  \url{http://edoc.ub.uni-muenchen.de/7/} 

\bibitem{Tum06d}
  Tumulka, R.:
  The `unromantic pictures' of quantum theory.
  \textit{J. Phys. A: Math. Theor.} \textbf{40}: 3245--3273 (2007).
  arXiv:quant-ph/0607124

\bibitem{Tum07}
  Tumulka, R.:
  Bohmian Mechanics at Space-Time Singularities. I. Timelike Singularities.
  arXiv:0708.0070

\bibitem{Val04}
  Valentini, A.:
  Black Holes, Information Loss, and Hidden Variables.
  arXiv:hep-th/0407032 

\end{thebibliography}
\end{document}